\documentstyle[12pt,aaspp4]{article}
\begin{document}
\def\emphasize#1{{\sl#1\/}}
\def\arg#1{{\it#1\/}}
\def\Mesz{M\'esz\'aros~}
\def\etal{{\it et~al.}}
\let\prog=\arg
\def\msun{M_\odot}
\def\simg{\mathrel{\hbox{\rlap{\lower.55ex \hbox {$\sim$}}
                   \kern-.3em \raise.4ex \hbox{$>$}}}}
\def\siml{\mathrel{\hbox{\rlap{\lower.55ex \hbox {$\sim$}}
                   \kern-.3em \raise.4ex \hbox{$<$}}}}
\def\edcomment#1{\iffalse\marginpar{\raggedright\sl#1\/}\else\relax\fi}
\marginparwidth 1.25in
\marginparsep .125in
\marginparpush .25in
\reversemarginpar

\title{Iron K Lines from Gamma Ray Bursts}

\author{T. R. Kallman$^1$, P. \Mesz$^2$ \&  M.J. Rees$^3$}

\begin{center}
\smallskip\noindent
$^1${NASA Goddard Space Flight Center, LHEA, Code 665, Greenbelt, MD 20771~~~~~~~~~~~~~~~}\\
$^2${Dpt. Astron. \& Astrophysics, Pennsylvania State University,
University Park, PA 16803}\\
\smallskip\noindent
$^3${Inst. of Astronomy, Cambridge University, Madingley Road, Cambridge
CB3 0HA, U.K.}\\
\end{center}

% \centerline{Oct. 23, 2001}
%\centerline{April 17, 2002}
\centerline{Oct. 29, 2002}

\begin{abstract}
We present models for reprocessing of an intense flux of X-rays and 
gamma rays expected in the vicinity of gamma ray burst sources.  
We consider the transfer and reprocessing of the energetic photons 
into observable features in the X-ray band, notably the K lines of iron.  
Our models are based on the assumption that the gas is sufficiently 
dense to allow the microphysical processes to be in a steady state, 
thus allowing efficient line emission with modest reprocessing mass 
and elemental abundances ranging from solar to moderately enriched.  
We show that the reprocessing is enhanced by down-Comptonization of 
photons whose energy would otherwise be too high to absorb on iron, 
and that pair production can have an  effect on enhancing the line 
production. Both ``distant" reprocessors such as supernova or wind 
remnants and ``nearby" reprocessors such as outer stellar envelopes
can reproduce the observed line fluxes with Fe abundances 30-100
times above solar, depending on the incidence angle. The high 
incidence angles required arise naturally only in nearby models, 
which for plausible values can reach Fe line to continuum ratios 
close to the reported values.

\end{abstract}

\section{Introduction}

The discovery of iron K line emission from the afterglows of 
cosmic gamma-ray burst sources (GRBs) provides a potentially 
important diagnostic of redshift as well as conditions in the 
burst environment.  The measured line intensities, together with 
distance estimates, constrain the total number of iron decays
needed to produce the line.  
A plausible assumption is that the line is formed by reprocessing of
continuum photons from the burst itself or its later outflow by
gas which is separate from the continuum-producing region, so that the
temperature and ionization are determined by those continuum photons.
If so, the time delay between the burst and the line detection provides
constraints on the timescales for recombination, level decay and light
travel time between the source of continuum source and reprocessor.
However, accurate calculations of the reprocessing of the burst continuum
spectrum into lines is potentially complicated owing to the possible
effects of radiative transfer, time dependent atomic processes affecting
line formation and the gas temperature.

The statistical quality of the observations so far are not sufficient 
to establish a unique model which can fit the available data.  
A key point is that the extremely large photon fluences near a GRB are 
likely to be sufficient to completely ionize the nearby gas in a time 
which is short compared with the duration of the burst.  
An important criterion for evaluating any model is the reprocessing 
efficiency, i.e. the ratio of the iron line fluence to the continuum 
fluence from the burst. 
Under a wide range of assumptions regarding the burst continuum flux and 
the reprocessor density the ionization parameter $\xi= L_{inc}/n r^2$ 
must exceed $\sim 10^3$ (see \S 2.2), where $r$ is the distance 
to the source and $n$ is the particle density. 
The other important observational constraint is
the time at which the line emission becomes most prominent, typically
several hours to a day for Fe emission features (e.g. Piro, \etal, 2000).
This places severe constraints on the reprocessing gas and has, so far, 
divided the models into distinct classes distinguished by assumptions 
regarding the density, thickness and composition of the reprocessing material.  

In the  ``distant reprocessor" scenario it is assumed that the gas 
illuminated by the burst and afterglow continuum is far enough for the 
time delay to be due to light travel time differences. In order to achieve
the required high ionization parameter, this involves reprocessing of the 
prompt burst emission, with luminosities of order $10^{50}$ erg/s.
The timescale for line emission per ion is not very short compared to the 
burst duration, so the reprocessing efficiencies are low if the iron abundance 
is solar, requiring therefore material which is highly enriched in iron in 
order to account for Fe line observations. Such conditions are expected to
occur in models with a distant shell or ring (e.g. Weth, et al, 2000, 
B\"ottcher, 2000, B\"ottcher \& Fryer, 2001, 
Ballantyne \& Ramirez-Ruiz 2001), possibly associated with 
supernova events (e.g. the supranova model of Vietri et al., 2001). 

On the other hand, ``nearby reprocessor" scenarios are capable of much greater 
reprocessing efficiencies, if the distances and gas densities are assumed to 
be comparable to those in extended stellar envelopes, or in dense, thick media 
similar to those in the reprocessors near accreting X-ray sources. 
In this scenario, iron enrichment is not required, and the timescales are 
sufficiently short that many Fe line photons are produced per Fe ion during
the afterglow phase of the burst. Incident luminosities of order $10^{47}$
erg/s are required at timescales comparable to a day, and radiative transfer 
effects are important.  Such conditions are expected, e.g., in the decaying 
jet model of Rees and \Mesz (2000), and in the jet plus bubble model of 
\Mesz and Rees (2001), which differ in the origin of the non-thermal component 
which is reprocessed by gas at distances comparable to outer stellar envelope.

In this  paper we present models which analyze these two generic families 
of models, with an added emphasis on the case of nearby reprocessors.  
Our aim is to provide an understanding of the spectrum formation physics and 
its dependence on the physical assumptions, rather than a specific model fit 
to the entire dynamics and light-curve histories of particular bursts.
In the following sections we present in turn the physics of line formation, 
our modelling technique, results, and a discussion of some implications.

\section{Background}

\subsection{Observations}

There are so far only a handful of observations of afterglows bright 
enough to allow unambiguous iron line detections (Piro et al., 1999, 2000; 
Yoshida et al., 1999).   Owing to the limited number of photons, and to 
uncertainties about the source redshift and the time evolution of the 
line, the line flux is the best constrained  observable quantity.  
Comparison with models requires conversion to luminosity, which is 
affected by the distance estimate.  As an example, we focus on the 
observation of GRB991216  by Piro et al. (2000) and distance estimate 
given those authors giving z=1.00$^+_-$0.02 and D=4.7 Gpc for $H_0=$75 
km s$^{-1}$ and $q_0$=0.5.  The maximum line luminosity was therefore 
$L_{line}\simeq 10^{53}$ photons s$^{-1}$ $\simeq 10^{45}$ erg s$^{-1}$, 
and a line fluence $\sim 10^{49}$erg. As discussed by Piro et al. (2000), 
Lazzatti et al. (1999), Vietri et al. (2000) and others the total mass 
of iron required to produce this is $\sim 50 M_\odot/n_{decays}$, 
where $n_{decays}$ is the total number of decays per iron nucleus.
This clearly demonstrates the tradeoff between $n_{decays}$ and the 
implied emitting mass of iron.  If $n_{decays} \gg 1$, then a moderate 
mass of iron is needed (e.g. Weth et al., 2000).  Piro et al. (2000) also 
report a possible detection of a feature attributed to the recombination 
continuum (RRC) of hydrogen-line iron, Fe XXVI.  This suggests that both 
features are emitted by recombination in highly ionized iron.

\subsection{Ionization Equilibrium}

If the line is emitted by reprocessing of burst continuum photons  then 
the value $n_{decays}$ cannot exceed the number of continuum photons 
absorbed per iron ion during the burst.  This limit is achieved when 
both the recombination  and photoionization timescales per ion are much 
less the duration of the burst.  
As an example, we adopt $L_{47}$ as the continuum luminosity in units of 
$10^{47}$ erg s$^{-1}$ and  $R_{13}$ for the distance from the source to 
the reprocessor in units of $10^{13}$ cm, and we assume a power law ionizing 
spectrum between 0.1 eV and 10 MeV with energy index -0.9.  This is a generic 
index which is is in the range of those observed in different afterglows in the 1-10 keV 
energy range relevant for Fe lines, (e.g. 
van Paradijs, Kouveliotou and Wijers 2000) and gives a substantial 
energy at $h\nu \geq 0.511$ MeV, which can contribute to pair formation. 
As will be seen, the pairs do not lead in the X-ray range to effects which
are greatly different than in the same calculations where pair-formation 
is artificially suppressed (Table 1). The results are less sensitive
to the properties of the spectrum above a few MeV than they are to changes
of the distances, densities, etc.
The recombination timescale from fully stripped into hydrogenic iron (case A) 
is approximately $t_{rec}\simeq 7 n_{11}^{-1} T_8^{0.74}$ sec where $n_{11}$ 
is the electron number density in units of $10^{11}$ cm$^{-3}$, and $T_8$ is 
the electron temperature in units of $10^8$K.  The photoionization 
timescale is $t_{PI}\simeq 2 \times 10^{-7} L_{47}^{-1} R_{13}^2$ s.
The preceding expression can be 
rewritten in terms of the ionization parameter $\xi=L/nR^2$ as  
$t_{PI}\simeq 2 \times 10^{3} \xi^{-1} n_{11}^{-1}$ s. This demonstrates 
that for parameter choices $10^{11}$ cm$^{-3} \siml n \siml 10^{17}$ cm$^{-3}$ 
and $\xi \geq 10^3$ erg cm s$^{-1}$ the recombination timescale is greater 
than the photoionization timescale, but that both are short compared with 
the burst duration ($t_{burst}\sim 10-100$s) or the delay between the burst 
and afterglow, delayed jet or bubble emergence ($t_{delay}\sim 10^5$ s).   
Therefore the gas can be regarded as being locally in ionization 
equilibrium:  the level of ionization will adjust itself such that the ratio 
of ionized (fully stripped) to non-ionized (i.e. hydrogenic, helium-like, 
etc.) iron will be equal to $t_{rec}/t_{PI}$,  and the value of this ratio 
is approximately $3 \times 10^{-3} T_8^{0.74} \xi$.  This demonstrates that 
the gas will be highly ionized in equilibrium for ionization parameters 
$\xi \geq 10^3$ which, as we will show, are most plausible for GRB 
reprocessors.

\subsection{Assumptions}

In the remainder of this paper we focus on gas densities in this range
$10^{11}$ cm$^{-3} \siml n \siml 10^{17}$ cm$^{-3}$.  Such densities are 
comparable to those expected in massive progenitor models of GRB, e.g. in
blobby ``distant" shells, 
or in ``nearby" envelope remnants illuminated by continuum from long-lasting 
jets or late emerging bubbles.  In addition, we assume that the temperature, 
ionization, and excitation conditions in the reprocessing gas are determined 
solely by reprocessing of continuum photons from the burst or its afterglow 
components.  We stress that in both of these scenarios the reprocessing gas is 
assumed to be moving with velocities $v \ll c$, i.e. essentially at rest.
That is, the reprocessing gas is physically separate from the gas responsible 
for the illuminating continuum photons in both models.  In the ``distant 
reprocessor" models ($R\geq 10^{15}$ cm the relativistic jet producing the 
continuum is assumed to be either inside the $v\ll c$ reprocessor shell (or 
the geometry is such that it does not matter if it is not) and the observed 
$\sim 10^5$s time lag between the burst and line detections arises from an 
$(r/c)(1-\cos(\theta)$ factor, where $\theta$ need not be a jet angle but may 
be the size of inhomegenities in the shell (Lazzati \etal 1999; Weth \etal 
2000; Piro \etal 2000; B\"ottcher \& Fryer 2001).  In the ``nearby reprocessor" 
models ($R\sim 10^{13}$ cm) the time lag of $\sim 10^5$ s arises because the 
outer parts of the stellar envelope (moving with $v\ll c$ are illuminated  by 
continuum from shocks caused by a weakened but long-lived jet (Rees \& \Mesz 2000) 
or by the late emergence of a bubble of waste heat (\Mesz \& Rees, 2001), in 
both cases at  $t\sim 10^5$ s.  
We also assume that dynamical effects of the burst on the reprocessor are 
unimportant.  Support for this comes from the fact that the sound crossing 
time for the most compact reprocessor we consider is $\geq 100$s, while
the microphysical timescales are much shorter. 

An added complication in comparing models with the data is the origin of 
the {\it observed} continuum in the vicinity of the line.  Inherent in our 
models is the assumption that the line photons are emitted nearly isotropically; 
some anisotropy results from radiative transfer effects as the line photons 
escape the reprocessor, which we assume is a geometrically thin and optically 
thick plane-parallel slab or thin shell (although we will also discuss 
alternatives to this in the next section).  
A portion of the continuum may be observed directly and be affected by 
relativistic beaming, and in the vicinity of the line this may differ 
from the continuum at the same energy which is incident on the reprocessor.  
On the other hand, the detectability of the line depends on the equivalent 
width measured with respect to these continuum photons.  
The above complications are highly geometry- and model dependent, and our
goal here is to investigate what can be said about such models based on
a minimum of assumptions and fairly general radiation physics, rather than 
trying to fit any particular model or observation.
In what follows we present model results and discussion using the quantity 
which is most closely related to the physics of the reprocessor:  the line 
luminosity.  We also discuss estimates for the observed line equivalent 
width based on simple assumptions regarding the continuum radiation field 
and the reprocessor geometry.

\subsection{Line Emission}

Iron line emission in photoionized gas occurs primarily by  recombination 
and  inner shell fluorescence.  The efficiency of fluorescence can be much  
greater than for recombination, but the ionization conditions required are 
less likely to be applicable to GRB reprocessors.

The efficiency of line emission by recombination is given by the product of 
the effective Fe XXVI L$\alpha$ recombination rate coefficient and the 
emission measure ionized by the burst continuum.  This quantity can be 
approximated as 

$$L_{rec}/L_{inc} \simeq 4 \pi N \alpha \varepsilon_{line} 
                                         y_{Fe} x_{FeXXVII}/\xi \eqno{(1)}$$

\noindent where  $L_{rec}$ is the line luminosity, $L_{inc}$ is the incident 
ionizing continuum luminosity, $N$ is the radial column density of the shell,
$\alpha\simeq 3.4 \times 10^{-13}$ cm$^3$s$^{-1}$ is the effective 
recombination rate coefficient for production of  the line, 
$\varepsilon_{line}$ is the line energy, 
$x_{FeXXVII}$ is the ionization fraction of fully stripped iron, 
$y_{Fe}$ is the iron abundance, and $\xi=L/n R^2$ is the ionization parameter
(eg.  Tarter Tucker and Salpeter, 1969).  This quantity is the ratio of the 
line luminosity radiated by an optically thin spherical shell to the 
luminosity of the central exciting source, and it enters the equation because of 
the fact that the recombination emission rate is proportional to gas density.
Inserting plausible numerical values gives

$$L_{rec}/L_{inc} \sim 1.4 N_{24} y_\odot x_{FeXXVII}/\xi \eqno{(2)}$$

\noindent where $N_{24}$ is the column in units of $10^{24}$ cm$^{-2}$.
The fractional abundance of highly ionized iron is negligible for ionization 
parameters $\xi\leq 10^3$ erg cm sec$^{-1}$, so the maximum fractional 
recombination luminosity attainable is $\sim 10^{-3}$.  

Inner shell fluorescence can occur if iron is not highly ionized, 
i.e. if the typical ion has  3 or more bound electrons. The luminosity is

$$L_{fl}/L_{inc} \simeq \omega_{fl} N \sigma_{PI} f_\varepsilon 
{{\Delta\varepsilon}\over{\varepsilon}} \varepsilon_{line} 
                                  y_{Fe} x_{\leq FeXXIV} \eqno{(3)}$$

\noindent where $\omega_{fl}$ is the fluorescence yield (0.34 for neutral 
iron), $\sigma_{PI}$ is the photoionization cross section at threshold, 
$f_\varepsilon$ is the normalized spectral function at the K threshold 
energy (see Kallman and McCray 1980 for a definition),
$\Delta\varepsilon/\varepsilon$ is a number of order unity which describes 
the fractional energy bandwidth contributing to the photoionization rate 
integral (for a power law with energy index $\alpha$ one has
$\Delta\varepsilon/\varepsilon=1/(3+\alpha)$), and 
$x_{\leq FeXXIV}$ is the ionization fraction of all iron 
ions with 3 or more electrons.  The rate for fluorescent 
line emission is proportional to the ionizing flux, so the efficiency is 
independent of $\xi$.  Inserting plausible numbers gives

$$L_{fl}/L_{inc} \sim 0.005  N_{24} y_{\odot} x_{\leq FeXXIV} \eqno{(4)}$$

\noindent for a single power law spectrum with an energy index of -0.9.

Another convenient measurement of line strength is the equivalent width 
measured with respect to the  continuum incident on the reprocessor, 
given by 

$$EW=(L_{line}/(L_{inc} f_\varepsilon) \simeq (L_{line}/L_{inc}) 
                              \varepsilon_{line} \kappa \eqno{(5)}$$

\noindent where $\kappa$ is a numerical factor depending on the shape of 
the incident continuum, $\kappa =ln(E_{max}/E_{min})\simeq$7 for our choice of power law with 
energy index -1 and where the continuum luminosity is measured between 13.6 eV and 13.6 keV.  
This shows that fluorescence lines can have equivalent 
widths $\sim$1 KeV, or fractional luminosities $\sim 10^{-2}$, but require 
low ionization parameter $\xi\leq 10^3$erg cm s$^{-1}$.  

For burst luminosities in the range 10$^{47}$ -- 10$^{52}$ erg s$^{-1}$ 
and distances $\leq 10^{15}$ cm from the source, gas densities 
$\geq 10^{15} \xi_3^{-1} L_{48} R_{15}^{-2}$ are implied if $\xi\geq 10^3$. 
This is at the limit of what we consider in the models which follow.
So, although fluorescence is included in all our models, it turns out to be 
generally less important than recombination, and the model equivalent widths 
(defined as in equation (5)) are less than the maximum attainable.

\section{Computational technique}
 
More accurate treatments of gamma-ray reprocessing and iron line emission 
require calculations of the ionization balance and electron kinetic 
temperature in the gas, along with the emission measure of gas which can 
emit lines.  Calculation of ionization balance and temperature is 
straightforward given the local mean  intensity of ionizing photons; 
calculating the transfer of these photons is complicated owing to scattering 
and attenuation by Compton scattering and photoelectric  absorption.  
Compton scattering can redistribute photons from high energies ($\simg$ 1 MeV) 
to energies where photoelectric absorption is important, and absorption  
depends on the local ionization balance.  In addition, high energy photons 
can produce $e^+ e^-$ pairs by collision with lower energy reprocessed 
photons, and these pairs can affect the ionization balance of iron by 
contributing to recombination, as well as contributing to Compton scattering 
of continuum photons.  An accurate treatment of all these process requires 
a numerical solution.

The problem of reprocessing of gamma rays and hard X-rays and iron line 
formation does not lend itself easily to most of the numerical techniques 
developed for treating either photoionized reprocessing or Comptonization.  
This is because Compton scattering is likely to be important if the column 
density of the reprocessor is large (eg. $\geq 10^{24}$ cm$^{-2}$) and 
because the scattering must be treated relativistically in order to 
accurately treat the gamma rays with energies $\simg$ 1 MeV.  Models 
developed for iron line formation in AGN accretion disks (see, for 
example, Nayakshin et al., 2000 and  Ballantyne et al. 2000) which 
utilize Fokker-Planck  or convolution methods for Comptonization are 
not highly accurate if these gamma rays are important.  At the same time, 
the reprocessing gas temperature must be $\sim 10^8$K or less, owing to 
limits on broadening of the observed lines, so that models 
developed for relativistic plasmas (e.g. Coppi and Blandford 1990) are also 
not directly applicable.

The reprocessing of gamma rays can affect the iron K line production 
in two ways.  First, down-Comptonization in a gas with mean electron energy 
which is small compared with the gamma ray energy will soften the spectrum 
in the interior of the reprocessor.  This can enhance the production of line 
photons owing to the increase in the cross section for iron photoionization 
at low energies.  In addition, incident gamma rays at energies greater than 
$\sim$ 1 MeV can produce pairs by $\gamma - \gamma$ collisions with reflected 
photons, if the center of mass energy is greater than the threshold for this 
process.  This can enhance the rate of recombination onto iron, 
and can also change the mean free path for Compton scattering.

The models presented here make use of the Monte-Carlo technique 
for treating the transfer of continuum photons.  This has the advantage that 
it allows for exact treatment of the relativistic rates and cross sections 
for Compton scattering and photon destruction, and that pair production can 
be incorporated in a straightforward way.  We use the Monte-Carlo code is 
described in Hua (1997), with modifications to allow for photoabsorption 
and pair production.  

In order to calculate line formation and photoelectric absorption 
it is necessary to combine the photon flux derived from the Monte-Carlo 
calculation with a calculation of the heating, ionization, and excitation 
of the gas.  To do so we use a photoionization equilibrium model (xstar; 
Kallman and Bautista 2000), using the ionizing flux from the Monte-Carlo 
calculation at each point in the cloud.  This calculation is iteratively 
repeated 3-5 times in order to self-consistently account for 
photoelectric absorption and Compton scattering.
Transfer of the line photons and thermal continuum photons escaping the cloud 
is calculated using the xstar escape probability formalism to calculate the 
local escaping flux, and then Comptonization of the line 
is calculated using an additional Monte-Carlo step.   

Input parameters are: ionization parameter $\xi$, the continuum spectral 
shape which we take to be  single power law with energy index -0.9 from 
0.1 eV to 10 MeV, gas density and elemental composition. 
The spectral index is chosen to be constant over the entire spectrum for simplicity, 
and because this value is representative of the observations in the 1-10 keV energy band.
Any departures from this value at energies below 1-2 keV which may be caused 
by absorption in the host or our own galaxy are unimportant for the purposes of modelling 
the iron K lines.
It is conventional (eg. Krolik McKee and Tarter, 1981; Kallman and Bautista, 2000) 
to define the ionization parameter in terms of the incident energy flux 
of photons integrated from 13.6 eV to 13.6 keV, and the gas (nucleus) number density, 
$\xi=4 \pi F/n$, where 
$F=\int_{13.6eV}^{13.6keV}{F_{\varepsilon}d\varepsilon}$.
The total incident photon number flux is 
$F^{(n)}_{inc}=\int_0^\infty{F_{\varepsilon}/\varepsilon}d\varepsilon$.
The chemical composition should be model- and assumption-dependent, 
hence in general a conventional (solar) choice is used, 
%(which also fits in well with nearby reprocessor scenarios) 
[H:He:C:O:Ne:Si:S:Fe]=[12:11:8.65:8.87:8.14:7.57:7.28:7.50]
(Morrison and McCammon 1981).
In Table 1 and \S \ref{sec:abundances} we explore the effect of
departures from solar abundances.

We perform the  Monte Carlo transfer calculation for a total number of 
photons $N_{tot}$ (100 photons per energy bin for each of 500 bins)
to get the number distribution of photons vs. depth and 
energy $N(\varepsilon, z)$.  At each scattering the cross sections for 
Compton scattering (fully relativistic in both photon and electron energy), 
photoelectric absorption, and pair production (described in more detail in 
the following section) are calculated.  The fate of the photon is determined
by calculating the path length for each process, also including escape, and 
taking the smallest. Then the number distribution of photons vs. depth and 
energy is is converted to a local energy flux using 
$F_{local}(\varepsilon,z)=F^{(n)}_{inc} N(\varepsilon, z)/N_{tot}$.
This then used to calculate the ionization balance and temperature throughout 
the slab using xstar.  Compton heating and cooling is calculated fully 
relativistically using the results of Guilbert (1986).  Both Monte-Carlo and 
photoionization steps are repeated a number of times ($\sim$4) to 
self-consistently calculate transfer, ionization , pair 
production/annihilation, etc.

\section{Model Results}

\subsection{Input and Output Parameters}
\label{sec:parameters}

The key issues we wish to address with these models are:

1) What is the penetration of the gamma rays and X-rays into the reprocessor, 
along with the down-Comptonization of gamma rays?  

2) What is the albedo of the reprocessor to gamma rays, and the spectrum of the 
reflected gammas?

3) What is the efficiency of iron line emission, and does it 
scale with $\xi$ as predicted by equations (2) and (4)?

To address these issues, we have run the models summarized in Table 1.  
These span a range of ionization parameter, and include models designed to 
test some of the assumptions described so far.  
The reprocessor distance range is 10$^{13}$ -- 10$^{16}$ cm, 
the gas densities are 10$^{11}$ -- 10$^{17}$ cm$^{-3}$,  and
the incident continuum luminosity range is 10$^{46}$ -- 10$^{52}$ 
erg s$^{-1}$. The smaller distance is comparable to a massive stellar
stellar envelope, while the densities range from those which might be
encountered in clumpy blobs of a $\sim 1 \msun$ shell at a light-day
distance, up to typical atmospheric or clumpy ejecta densities at
massive stellar envelope distances.  The high values of incident luminosity 
$L_{inc}$ are characteristic of the earlier epochs (minutes), while the 
lower ones are characteristic of later epochs (several hours to days).
The ionizing spectrum is held constant throughout.  We have also explored 
the dependence of models on the iron abundance, the angle of incidence on the
model slab, and on computational assumptions such as the treatment of pairs and 
Comptonization.  

Table 1 presents the results of our model calculation: ionization parameter 
and iron line strength, Fe XXVI L$\alpha$ alone (in the column 
labeled L$_{Fe26}$) and total Fe K in the 6.4 -- 6.9 keV range 
(in the column labeled L$_{Fe}$).  These 
line luminosities correspond to the total emitted line luminosity, without 
attempting to distinguish the un-Comptonized fraction.  
Of more interest are the observable quantities: the line luminosities which 
escape unscattered and their strength relative to the continuum.  Table 1 lists 
the luminosity of the iron line which escapes unscattered,  L$_{FeUs}$, and two 
different measures of the line reprocessing efficiency:  the ratio of the 
unscattered line to the reflected continuum luminosity  L$_{FeUs}$/L$_{ref}$, 
and the equivalent width.  For our purposes the reflected luminosity is 
integrated over the 1 -- 10 keV energy band and calculated according to 
L$_{ref}$=albedo $\times$ L$_x$, and both the albedo and L$_x$ are given in the 
table.  The equivalent width is calculated as the ratio of the integrated 
residual flux in the unscattered components of the iron line
to the averaged inter-line continuum in the 6.5 -- 7.1 keV energy range.
These quantities are all calculated in the source rest-frame.

\begin{table}
\begin{tabular}{cccccccccccccc}
Model &n&r$_{in}$&$\xi$&$\mu$&Fe&E$_{cut}$&L$_{Fe26}$& L$_{Fe}$&L$_{FeUs}$&Albdo&L$_x$&EW
&$\frac{{\rm L}_{FeUs}}{{\rm L}_{ref}}$\\
1             &17&13&3&1   &  1&6.3&44.11&44.28&42.59&0.71&45.5&0.001&-2.76\\
2             &17&13&4&1   &  1&6.3&44.37&44.45&42.27&0.82&46.5&0.003&-4.14\\
2np           &17&13&4&1   &  1&6.3&44.37&44.57&42.46&0.83&46.5&0.003&-3.96\\
2nc           &17&13&4&1   &  1&6.3&43.15&43.40&41.73&    &46.7&     &    \\
2'            &17&13&4&0.2 &  1&6.3&43.95&44.18&43.6 &0.97&46.5&0.048&-2.89\\
2''           &17&13&4&0.05&  1&6.3&43.1 &43.6 &43.7 &0.97&46.5&0.014&-2.79\\
2x30          &17&13&4&1   & 30&6.3&45.59&45.71&43.49&0.74&46.5&0.009&-2.88\\
2x100         &17&13&4&1   &100&6.3&45.96&46.2 &43.56&0.70&46.5&0.040&-2.79\\
2''x30        &17&13&4&0.05& 30&6.3&43.94&44.53&44.14&0.83&46.5&0.770&-2.28\\
2''x100       &17&13&4&0.05&100&6.3&44.1 &44.98&44.7 &0.79&46.5&1.280&-1.70\\
2''x100s      &17&13&4&0.05&100&5.0&44.54&44.66&45.54&0.49&46.5&1.440&-1.65\\
3             &17&13&5&1   &  1&6.3&44.77&44.99&42.34&0.90&47.5&0.001&-5.11\\
4             &17&13&7&1   &  1&6.3&45.51&45.64&42.25&0.86&49.5&0.019&-7.18\\
5np           &17&13&9&1   &  1&6.3&45.71&45.81&42.13&0.86&51.5&0.025&-9.30\\
5             &17&13&9&1   &  1&6.3&45.69&45.79&42.11&0.87&51.5&0.025&-9.33\\
-      &- &- &-&-   &-  &-    &-    &-    &-   &-   &-    \\

6             &11&16&4&1   &  1&6.3&44.29&44.55&42.34&0.81&46.5&0.006&-4.07\\
6x100         &11&16&4&1   &100&6.3&45.81&46.44&43.51&0.70&46.5&0.011&-2.84\\
6x100s        &11&16&4&1   &100&5.0&46.11&46.47&45.30&0.90&46.5&0.061&-1.15\\
6ni           &11&16&4&1   &  1&6.3&     &44.88&     &0.80&46.7&     &     \\
6'            &11&16&4&0.2 &  1&6.3&43.7 &44.1 &43.6 &0.95&46.5&0.220&-2.88\\
6''           &11&16&4&0.05&  1&6.3&42.88&44.09&43.88&0.92&46.5&0.970&-2.58\\
6''x30        &11&16&4&0.05& 30&6.3&43.6 &45.52&45.4 &0.83&46.5&1.030&-1.02\\
7np           &11&16&9&1   &  1&6.3&45.91&46.04&42.27&0.87&51.5&0.020&-9.17\\
7             &11&16&9&1   &  1&6.3&45.91&46.04&42.28&0.86&51.5&0.020&-9.15\\
\end{tabular}
\caption{\footnotesize Results for ``nearby" models 1-5 and ``distant" models 6-7.
Parameters are the reprocessor density $n$ (cm$^{-3}$), distance $r_{in}$ (cm),
ionization parameter $\xi$ (based on the incident continuum luminosity
in the 13.6 eV -13.6 keV range), the cosine $\mu$ of the incidence angle
relative to the surface normal, the Fe abundance in solar units, and the incident
power law cut-off in eV.
In nearby models (1-5) time delay effects are unimportant, and the line
luminosities L$_{Fe26}$ and L$_{Fe}$ correspond to H-like and total Fe,
while the L$_{FeUs}$ column is the unscattered Fe line luminosity.
The last four columns are the 1-10 keV X-ray albedo, incident X-ray luminosity, the
line equivalent width in keV, and the unscattered Fe line to 1-10 keV reflected
continuum ratio. All values are in the rest frame. Quantities are logarithmic,
except for Fe abundances, cosines $\mu$, albedo and equivalent width.
The equivalent width is calculated as the ratio of the integrated flux in the
narrow components of the iron line
to the averaged inter-line continuum in the 6.5 -- 7.1 keV energy range.
Conversion into line emission is much less efficient for high $\xi$, because
the emission saturates at a level determined by the recombination rate,
which depends primarily on $n_p$.
For models 6 and 7 the table gives the instantaneous specific line
luminosities; the observed line-strengths would be reduced by a
time-delay smearing factor in the integration over the large reprocessing
shell. The luminosity in model 6 is typical of an afterglow after one day,
so time-smearing leads to only a modest reduction. On the other hand, the
luminosity of model 7 would be relevant for the first $\sim$ 10 seconds, and
time-smearing would lead to a more significant reduction in the observed line
luminosity (\S \ref{sec:variab}).
{\it Comparison models:} 1) np= no pairs; 2) nc = no comptonization;
3) ni= Nickel (Fe=0, Ni=20, see text). The model 6ni line luminosity is for
the blend of nickel K lines; 4) s=soft (100 keV) incident power-law cut-off.}
\end{table}

\subsection{Total Emitted Line Luminosity}
\label{sec:linelum}

In Figure 1 we plot the line luminosities from table 1 
as a function of ionization parameter for the 
nearby models.  This shows the behavior predicted by equations (1) and (2), ie. 
that the line reprocessing efficiency decreases approximately inversely with 
increasing ionization parameter.  For the ``distant" models at R=10$^{16}$ cm
we find that the $L_{line}/L_{inc}\propto \xi^{-0.75}$.  Comparison of 
equation  (2) with the line formation efficiencies derived from the total line
luminosities in the table shows a 
difference of a factor $\sim$ 10 -- 50.  This is due  to the influence of 
Comptonization, which increases the flux of photons available to ionize 
iron and hence the line emitting volume.  A further illustration is 
provided by the results of model 2nc, which was calculated with the same 
parameters as model 2, but using simple single stream exponential attenuation 
of the incident photons rather than using the Monte Carlo Comptonization 
calculation.  The efficiency of model 2nc is very nearly the same as 
that predicted by equation (2).

Question (1) can be addressed by comparing models 2 and 2nc.  The 
effect of an accurate transfer treatment is to allow penetration of 
gamma rays due to the enhanced forward scattering probability of the KN 
cross section, and also to allow down-Comptonization of these penetrating 
photons at large column depths in the slab.  For the estimates in the 
previous section we took $\tau_{Th}\simeq 1$ for the Thompson depth of 
the ionized part of the slab, but the Monte Carlo results show that photons 
penetrate to much greater depths.  In Figure 2a we show a contour plot of 
photon intensity vs. energy and depth for model 2, which shows the incident 
radiation field is not depleted until $\tau_{Th}\simeq 10$.  This accounts 
for the  greater line intensity in model 2 than in model 2nc.

Figures 2a-d give more details of the spatial distributions of various 
physical quantities in model 2 for normal incidence.
Figure 2a shows a contour plot of the ratio of photon mean intensity in the 
interior of the model to that at the surface as a function of energy and 
optical depth.  The contour spacing is a factor of 1.6 in this figure, 
and dashed or solid contours indicate regions where the ratio is less than 
or greater than unity, respectively.  This shows that photons below 
$\sim$ 100 keV penetrate to $\tau_{Thompson} \sim 10$ for this choice of 
parameters before the intensity falls below 0.1 of the surface value.  
The mean intensity increases with increasing depth to a maximum at 
$\tau_{Thompson} \sim$ 3.  The flux would be zero in a pure scattering 
slab, but the effects of photoabsorption and reemission which shifts 
photons into the UV results in a non-zero net flux.  
%The island near $\tau \sim 1$ and energy 0.1-a few keV is expected 
%due to the effects of down-Componization,
%
Figure 2b shows the electron temperature vs. depth.  Near the surface
the gas approaches the Compton temperature, $T_{IC}$, which in this case 
is $\simeq 3 \times 10^8$ K.  The blip near $10^5$ K is a common feature in 
thermal equilibrium curves, being related to non-linearities in the heating 
and cooling from intermediate mass elements such as oxygen.  Collisional cooling 
has a temperature dependence $\exp(-\hbox{const}/T)/\sqrt(T)$, and 
the local maxima in this function can lead to such bumps. 
Figure 2c shows the distribution of emissivity with depth in model 2.
The various curves correspond to the components of the iron line from 
the hydrogen-like (dotted), helium-like (dot-dashed) and lower 
(dashed) ion stages.  This reflects the dominance of recombination onto 
hydrogen-like iron in this model, owing to its large ionization parameter. 

\subsection{The Effects of Comptonization on the Line Escaping Spectrum}
\label{sec:compton}

The mean wavelength shift for iron K line photons per Compton scattering in 
the hot, ionized part of the reprocessor is $\Delta\varepsilon \simeq$ 0.46 KeV.  
This is sufficient to smear line photons beyond recognition, so as a practical 
matter we expect that only unscattered photons will be recognizable as being 
associated with a line.  We have modelled the effects of Comptonization on the 
line and thermally emitted continuum photons created in the models using the 
procedure described in the previous section.  Figure 3a shows the spectrum in 
the 5-10 keV energy band for model 2 at normal incidence.  The dashed curve shows 
the total emitted spectrum integrated over the slab, including the Lyman series 
lines and Lyman continuum emission of Fe XXVI, and the 1 -- n lines and 
recombination emission of Fe XXV.  This model also emits  Lyman lines and 
recombination continua from O VIII and lines from highly ionized Si and S 
(not shown).  The lowest of the solid curves shows the Comptonized line and 
thermally emitted continuum spectrum escaping the cloud, calculated 
self-consistently using a Monte-Carlo treatment.  Statistical uncertainties 
associated with the Monte Carlo treatment of 
scattering in this figure are small; we emit 1000
photons in each of the energy bins in our model; the few apparent 
gaps in the injected spectrum are due to slight 
mismatches in mapping between the energy grids used for the 
Monte Carlo and the xstar part of the calculation.
For the purpose of treating the line escape from the cloud we 
have added the process of line resonant scattering to the Monte-Carlo 
calculation, so that we account for the enhancement to the photon path length 
and the probability of Compton scattering for resonance line 
photons such as Fe XXVI L$\alpha$.  In doing so, we assume that each line 
scattering event is treated according to complete redistribution in the line 
Doppler core, and completely coherently in the line wings.  In practice, the 
latter is unimportant, since  Fe XXVI L$\alpha$ has maximum depth $\sim$1000, 
and the damping parameter is $\simeq 10^{-4}$.  The results of the calculation
of the luminosity in the unscattered core of the Fe XXVI L$\alpha$ line, for 
all our models, is also given in the column of the Table labeled $L_{FeUs}$.  
The upper solid curve, plotted with coarser binning, is the reflected Comptonized 
continuum for this model.

The results of Figure 3a for normal incidence show that the fraction of the 
line photons escaping unscattered is $\sim$0.01; most of the luminosity
escapes as a broad comptonized continuum in the vicinity of the line.   
The narrow core of the 6.97 keV line  has an equivalent width
$EW \simeq 3 eV$, and a fractional luminosity 
$L_{FeUs}/L_{ref}\simeq 7.2 \times 10^{-5}$ measured relative to 
the total scattered flux integrated over the 1 -- 10 keV energy band.
This figure, as well as Figure 2c, shows that although 
various components of the line are emitted, the components at energies 
6.7 keV (He-like) and below are more Comptonized than the component at 
6.97 keV (H-like). This is due to the fact that the higher energy (H-like)
component is emitted in the shallower, more highly ionized gas closer to 
the slab boundary, and therefore traverses a smaller depth as it escapes.  
The recombination continua are apparent in the emitted spectrum but are
unrecognizable in the scattered escaping spectrum.  
Resonance scattering does not affect the photons emitted in the higher Lyman 
series lines or the  recombination continuum as much as the L$\alpha$ line, 
so the ratio of these lines to the L$\alpha$ analog line exceeds the 
recombination value in our simulations.  

The difference in the luminosity escaping unscattered and the emitted luminosity 
is displayed for all models in table 1, and for the nearby models is plotted in 
figure 1.  This shows that the effects of Comptonization of the escaping line are 
greater at high $\xi$.  This is due to the relative importance of scattering and 
photoelectric absorption as a function of ionization parameter.  At high $\xi$ 
photoelectric absorption is reduced in importance, so that incident photons 
penetrate to greater Compton depths before being absorbed and reemitted as 
iron line photons.

The effects of Comptonization on the escaping line profile are reduced if the 
line resonance scattering optical depth scale is changed.  This might occur if, 
for example, the cloud had a large internal velocity dispersion 
$\sim$3000 km s$^{-1}$.  Numerical experiments show that this affects the 
unscattered line luminosity by  a factor $\leq$2, reflecting the fact that 
the regions of large line depth are also regions of large continuum 
Thomson depth, and photons emitted in these regions are likely to be 
Comptonized even if they escape without resonance scattering.  

\subsection{Incidence Angle and Abundance Dependence of the Lines}
\label{sec:incidence}

The depth scale for scattering of all photons is affected if the incident photons 
hit the slab at a non-normal angle.  The effects of this are displayed in Table 1,
Figure 1, and Figure 3b.  As shown in Table 1 and Figure 1, if the incident angle 
is 70$^o$ ($\mu=0.2$) the unscattered line luminosity is increased by a factor 
10-30 compared with normal incidence.  If the incident angle is increased to 
87$^o$ ($\mu=0.05$) then the effects of Comptonization on the luminosity of the 
narrow line core are negligible; the escaping unscattered line luminosity is 
comparable to the total emitted. 

This is illustrated in Figure 3b, for model 2'', which has an incidence angle of 
87$^o$, for comparison with the normal incidence case of model 2 in Figure 3a. 
Also notable in this figure is the fact that the fraction of the unscattered 
He-like line, near 6.7 keV, is greater than for the H-like line at 6.97 keV.  
This is significantly different from the results for normal incidence, 
for which the unscattered fraction of the He-like line is negligible.  
This is due to the differing scattering behaviors of the two lines: the H-like 
line is subject to resonance scattering while the forbidden and intercombination 
components of the He-like line are not, but the H-like line is emitted closer 
to the illuminated surface.  At normal incidence the disparity in depths of 
emission is more important than the difference between resonance and 
non-resonance scattering, while at 87$^o$ the converse is true.
The net effect of non-normal incidence is to reduce the Thomson depth of the 
hot ($\sim 10^8$K) scattering layer of the model clouds.  
Evidence for this in the results of model 2'' and other non-normal incidence models is the 
`shoulder' below the 6.7 keV line in the scattered spectrum.  This is due to 
the fact that in all models a significant fraction of the 6.4 -- 6.7 keV line photons penetrate 
into the cold part of the cloud and scatter there.  In this region the Compton energy 
shift per scattering is always to lower energies, $\simeq$0.08 keV, thus reducing a low 
energy shoulder on the line.  In the non-normal incidence models many of these 
photons escape owing to the reduced optical depth of the hot cloud layer.
Although this shoulder is narrow enough that it might be interpreted as being part of  the 
line by low resolution instruments, we do not include it in our accounting for escaping unscattered 
photons.  We do include it in the continuum accounting, and affects the equivalent widths we 
derive as discussed in the following section.

In most of our models the iron line emission is dominated by recombination, 
so that the total line luminosity is expected to scale with the increase in 
the iron abundance, and at least for the nearby reprocessors this scaling is
geometry independent. An increased iron abundance results in enhanced 
cooling and lower equilibrium temperature, which in turn increases the 
recombination rate coefficient, and the penetration and escape depth are
also affected in a non-linear manner. We have carried out experiments using 
model 1, the lowest ionization parameter case, indicating that the total 
emitted iron luminosity scales approximately as the square root of the iron 
over-abundance for enhancements of up to a factor of $\sim 10^2$ over solar 
values.  The escaping unscattered component of the Fe XXVI L$\alpha$ line 
scales somewhat more slowly  with the abundance due 
to the effect of resonance line trapping; e.g. for normal incidence 
reprocessing the unscattered line scaling with abundance is 
d log(L$_{Fe}^{(us)}$)/d log(Fe abundance) $\sim$ 0.4.

For two models chosen as representative of the nearby and distant cases,
models 2 and 6, we have performed calculations where the Fe abundance is 
30 and 100 times the solar value.  Figure 4 show the results for two nearby 
models, model 2''x30 (85$^o$ incidence angle with Fe 30$\times$solar, 
panel a), and model 2''x100 ($85^o$ and Fe 100$\times$ solar, panel b).
A comparison with the similar inclination but solar abundance model 2'' in 
Figure 3b shows that increasing the iron abundance results in an increase 
in the line luminosity, both emitted and escaping, 
by factors which are approximately consistent with the the square root 
scaling described above; model 2''x100 gives an enhancement in the escaping 
unscattered luminosity by a factor 10 over that for solar model 2''.
Compared to a model with normal incidence and solar abundance, model 2"x100
has an unscattered Fe line luminosity a factor $\gtrsim 270$ times larger,
$L_{FeUs}= 5 \times 10^{44}$ erg s$^{-1}$, a fractional line luminosity 
$L_{FeUs}/L_{ref}=0.02$, and an equivalent width EW=1.28 keV.

Figure 5 shows the results for two distant cases, 
model 6''x30 ($85^o$ incidence with Fe 30$\times$solar, panel a), and
model 6x100 (normal incidence with Fe 100$\times$solar). 
Comparison of the nearby model 2''x30 (Figure 4a) and the comparable distant 
model 6''x30 (Figure 5a) shows a strong 6.4 keV fluorescence component in 
the distant model which is not present in the nearby models. The Fe XXVI 
and Fe XXV components are of comparable strength in the two models. The 
presence of the fluorescence component is due to penetration of ionizing 
photons into the partially ionized zone of the distant model.  This does 
not show up in the nearby models due to the fact that the size of the 
ionization fronts in photoionized models scale proportional to $\sqrt{Ln}$  
(Kallman and McCray, 1982; McCray Wright \& Hatchett, 1978), 
so that in the nearby models fewer photons capable of photoionizing 
iron penetrate into the partially ionized zone of the model slab.  
The large incident angle then implies that many of these photons can escape 
un-Comptonized; the normal incidence distant models produce these photons at 
greater depths, such that Comptonization smears them as they escape. The  
unscattered Fe line flux from model 6''x30 is $L_{FeUs} \sim 2.5\times 10^{45}$ 
erg/s,  a factor $\sim 5$ larger than for model 2''x100 shown in Figure 4b.  
The fractional line luminosity from this model 6''x30 is $\sim$ 0.1 and
the equivalent width is 1.03 keV.  This is an example of a model where the 
continuum near the unscattered line is dominated by the Compton shoulder, 
while the 1-10 keV continuum is dominated by the scattered continuum.  The result 
is that the line/continuum ratio is the largest of all the models in table 1, 
but the equivalent width is not.

The distant models do not show as large line/continuum ratios and equivalent widths
when low incidence angles are assumed. This is seen in Fig. 5b for model
6x100, for normal incidence and Fe abundance 100$\times$ solar,
giving a line/continuum ratio of 0.0015 and equivalent width 0.011 keV.

\subsection{Comparison of Model Properties and Dependences}
\label{sec:comparison}

Comparing Figures 3, 4 and 5, we see that a high inclination increases 
substantially the escaping unscattered line fluxes, as does increasing the Fe abundance. 
For similar ionization parameters, chemical abundances and normal incidence, 
the distant models appear to produce only marginally larger line fluxes 
than nearby ones. This approximate parity remains as one increases the 
chemical abundances.  However, for large incidence angles, the distant models 
appear to produce larger line fluxes than nearby ones, by factors up to 
$\sim 10$, other factors being similar. This is seen in the line to 
continuum ratios and equivalent widths of Table 1. For the same inclination
angle and overabundance, the line/continuum ratio is a factor 20 larger
in model 6''x30 than in model 2''x30.   Model 6''x30 (Figure 5a) gives a 
line/continuum ratio of 0.1, which is also a factor 5 larger than 
Model 2"x100 (Figure 4b), which gives 0.02. This reflects the fact that at
larger incidence angles the line arises from shallower depths, and the lower
densities of the distant models lead to more penetration of the iron line 
photons into the partially ionized zone of the slab.  
The disparity between nearby 
and distant models is greatly reduced  when the line equivalent width is 
considered.  The equivalent width of all the non-normal incidence models with 
enhanced iron are comparable to within $\simeq 50\%$, reflecting the 
influence of the Compton shoulder on the continuum near the line and 
variations in the shape of this feature from one model to another.

The absolute values of the line/continuum ratio 
and equivalent widths in the normal incidence case
are not only similar for distant and nearby models, but are also much lower 
for the normal incidence angle cases (Table 1). This is also seen from a 
comparison of Figure 3a (model 2, normal, solar) and Figure 3b (model 2", 
85 degrees, solar): the line is much stronger in the inclined case. This is
also the case even if one boosts the abundance in the normal incidence case,
as in Figure 5b (model 6x100, normal, 100xsolar), compared to Figure 5a 
(model 6"x30, 85 degree, 30xsolar); even though the latter has lower 
abundance, its unscattered line is stronger than in the higher abundance, 
normal incidence case.

We have also examined the effect on our models of reducing the maximum photon energy 
in the illuminating spectrum from 10 MeV to 100 keV for models 2''x100 and 6x100.  
This choice of cutoff is similar to that used by Ballantyne et al. (2001).
The ionization parameter, and therefore the incident flux in the 13.6 eV -- 13.6 keV band,
is the same as in the other versions of models 2 and 6.  This has the effect of 
eliminating pair production, and of lowering the Compton temperature to 3 $\times 10^7$ K.
This has the effect of increasing the recombination rate, and thereby the line 
luminosities.  Since the 13.6 eV -- 13.6 keV flux is held constant, the total incident flux 
is lower in the 100 keV cutoff models, and therefore so is the total energy 
deposited in the slab.  This has the effect of reducing the energy in the emitted line, 
particularly in the deepest parts of the cloud.  The escaping line luminosity is 
increased by $\sim 20 \%$ for model 2''x100, since in this model much of the line 
comes from deep in the cloud and the two effects act oppositely.  In model 6x100 most of the 
escaping unscattered line comes from the recombination region, so the 100 keV cutoff model 
significantly enhances the escaping line.

Interpretation of these results in 
terms of observations requires the introduction of additional assumptions.  
Although the most straightforward observational quantity that can be derived 
from the models is the total line flux emitted by the cloud, observations 
of the line are affected by Compton broadening of the line and by the 
statistical significance of the line relative to the adjacent continuum.  
The line and continuum have differing dependence on  reprocessing:  the 
line must be reprocessed, while the continuum may include direct 
(unreprocessed) radiation.  
Moreover, we can envision various geometrical configurations for our model 
reprocessors even within the assumptions of a time-steady unbeamed continuum 
source.  If the reprocessor has a covering fraction relative to the source 
less than unity, then the observed continuum near the line would likely be
dominated by photons from the source.  If the photons from the source are 
not directly observable, either due to time delays or due to beaming away 
from us, then the observed continuum near the line would be entirely due 
to Compton reflection and emission from the reprocessor. This simplified 
continuum scenario is what we consider in figures 3 and 4. 
Calculations of the line luminosity presented so far have illustrated the 
importance of Compton down-scattered gamma rays on the spectrum emitted in the 
cloud interior.  Since this process occurs primarily at large depths in the 
reprocessor, line photons must traverse a corresponding column to escape, 
and scattering during this process broadens the line.  
Although we consider these 
results to be an accurate prediction of the spectra corresponding to our assumed 
choice of parameters, they are likely to be quite  sensitive to our assumptions 
regarding the reprocessor geometry.  Clearly, the assumption of normal incidence 
onto a plane-parallel slab will produce the 
lowest fraction of escaping unscattered line photons.  Other 
geometries, such as non-normal illumination or non-plane-parallel reprocessors, 
will produce a greater fraction of 
photons created at large depths which will escape unscattered.  Therefore, we 
consider it likely that real reprocessors will produce lines with luminosities  
in a range between the unscattered luminosities and the total emitted 
luminosities (for normal incidence) given in the table.  

Our  results differ from  the calculations of 
McLaughlin et al. (2001) who considered Comptonization in a funnel 
geometry in that the reprocessors considered here have a Compton temperature 
$\simeq 10^8$K, so that the mean energy shift per scattering is large.  Thus 
we do not predict easily identifiable spectral features associated with once-
or twice-scattered photons, even for photons emitted at small depths.
We have not explored different assumptions about the incident continuum 
shape, which could lead to reduced Compton temperature.  We note, however,
that it is unlikely that the temperature in the line emitting region 
will be low enough (i.e. $\leq 10^6$K) that the effects of thermal 
broadening in the Compton escape will be negligible.

Our results shown in figure 3b are similar to those calculated by Ballantyne 
and Ramirez-Ruiz (2001) in the effects of Comptonization on the lines from 
H-like and He-like Fe.  However, the difference between figures 3a and 3b 
illustrate the dependence of this result on the assumed geometry; at normal 
incidence the behavior is qualitatively different.  Our models differ from 
those of Ballantyne and Ramirez-Ruiz (2001) in the choice of high energy 
cutoff for the illuminating continuum, and therefore in the Compton temperature.  
Our model continua extend to 10MeV and have a Compton temperature $\simeq 3 
\times 10^8$ K.  A consequence of this is a lower recombination rate 
coefficient, and correspondingly lower reprocessing efficiency.  

\subsection{Pair Production}
\label{sec:pairs}

Pair production occurs due to $\gamma , \gamma$ collisions between incident 
and reflected photons.  We use rate coefficients taken from Coppi and 
Blandford (1990), equation (4.6):

$$R_{prod} \simeq c \sigma_{Th} (F/(c\varepsilon_{ave}))^2 f_1 f_2 \eqno{(6)}$$

\noindent where $\varepsilon_{ave}$ is the average photon energy, and 
$f_1$ and $f_2$ are factors less than unity describing the penetration of 
gamma rays and the albedo for upward photons, respectively.
The rate of destruction by annihilation is approximately 

$$R_{dest} \simeq {3\over 8}  c \sigma_{Th} n_e n_p \eqno{(6)}$$

\noindent Equating these gives an equilibrium pair density relative to 
protons of 

$$n_e/n \simeq {\xi\over{c\varepsilon_{ave}}} \sqrt{ 8 f_1 f_2/3} \eqno{(7)}$$

\noindent which is $n_e/n \simeq 10^5 \sqrt{ 8 f_1 f_2/3}$ for $\xi = 10^4$, 
being proportional to $\xi$.  Thus pair production can significantly enhance 
the total electron number density if the incident spectrum has a significant 
flux above $\sim$ 1 MeV, and if the reprocessor albedo is not negligible.  
The effect of pairs on the iron line will be twofold: to enhance the iron 
recombination rate and thereby the line luminosity, and to decrease the mean 
free path of photons to Compton scattering.  The Thompson depth in a 
pair-dominated cloud is proportional to $n_e R \propto \xi n R \propto L/R$, 
the compactness parameter (e.g. Coppi and Blandford 1990)

In order to evaluate quantitatively the  effects of pairs we have included 
a calculation of the pair formation rate and of the equilibrium pair density 
self-consistently in all our models, using the following procedure:  In our 
iterative procedure we initially set the continuum opacity and upward flux 
of gammas to be zero. As part of the  Monte Carlo transfer calculation 
we calculate the number of pairs produced as a function of depth $N_{pairs}(z)$, 
and the number of upward photons vs. depth and energy, $N_{up}(\varepsilon, z)$. 
Pair production is calculated using the rate coefficient $R(x)$ from 
Coppi and Blandford equation 4.6 together with the number of upward photons 
vs. depth.  The cross section for pair production is given by: 
$\sigma_{pair~prod}=\max(R(x) F^{(n)}(\varepsilon,z))/c$, and the maximum is 
taken over all upward photon energies, and  $F^{(n)}(\varepsilon,z)$ is the local 
upward photon number flux as a function of energy $\varepsilon$.  The value of
$F^{(n)}(\varepsilon,z)$ is calculated from the number of upward photons 
$N_{up}(\varepsilon, z)$ by  
$F^{(n)}(\varepsilon,z)= F^{(n)}_{inc} N_{up}(\varepsilon, z)/N_{tot}$.  
This step is  repeated a number of times ($\sim$10) to self-consistently 
calculate the pair production and upward flux of gammas.  Pair production is 
included in the xstar calculation by converting the number of pairs created 
at each depth to a rate by 
$R_{pair~prod}(z)= F^{(n)}_{inc} N_{pairs}(z)/N_{tot}$.
Destruction (annihilation) is calculated using the rate given in Coppi and 
Blandford (3.7).  The equilibrium pair density is added to the ordinary 
free electron density and allowed to contribute to recombination, line 
formation, etc.  This is equivalent to assuming that the pairs thermalize 
before annihilating.  This can be justified by noting that the timescale for 
slowing down a fast ($\sim$ 1 MeV) particle in a fully ionized gas is 
approximately $10^3 n_{12}^{-1}$ s, while the $e^+ e^-$ annihilation 
timescale is longer by a factor $\sim$a few (Bussard et al., 1979).
Both Monte-Carlo and photoionization steps are repeated a number of times 
($\sim$4) to self-consistently calculate transfer, ionization , 
pair production/annihilation, etc.  We also implement a  self-consistent 
updating of the optical depth scale when iterating between the xstar and 
Monte-Carlo parts of the problem.  This is done by performing the 
Monte-Carlo in optical depth space, so the only density dependent quantity 
is the ratio of pair cross section to scattering cross section for each 
flight. The pair production rate is found to converge to within 
$\simeq$10$\%$ after 3 iterations between the Compton and xstar parts of 
the calculation if the reflected flux is initially assumed to be zero.  

The effects of pairs are shown in the table by comparing models 2, 5 and 7 
with the corresponding models which are identical but which have the 
effects of pairs turned off (models 2np, 5np and 7np respectively).  
At each point in the model the local effect of pairs is to increase the 
recombination rate and thereby to decrease the level of ionization of 
the gas, increasing the density of ions such as Fe XXVI and Fe XXV.  
Since the photoionization heating rate increases proportional to these 
abundances, the effect is to increase the gas temperature.  This, in turn, 
increases the optical depth to photoabsorption by highly ionized iron, 
thereby reducing the photoionization heating rate deeper in the cloud.  
Pairs cause the temperature to be greater at small depth, and lower at 
large depth than would otherwise be the case.  Since the iron line 
emissivity is generally a decreasing function of temperature for 
photoionized models, the two regions will have competing effects on the 
total line luminosity.    The results of the Table  show that the line 
luminosity is unchanged by pairs for models 2 and 7, and is slightly 
decreased by pairs for model 5.  This difference between models can be 
attributed to the greater compactness of model 5 compared with either model 
2 or 7.  The unscattered iron line luminosity is affected more by the 
neglect of pairs than is the emitted line luminosity, reflecting the fact 
that pairs affect the optical depth scale more than the temperature
distribution.  
An additional effect of pairs, for reprocessors with densities less than we 
consider here, is to reduce the Compton mean free path relative to the 
cloud size.  This can allow clouds with low (proton) column densities 
to be Thompson thick.

\section{Effects of Continuum Variability and Time Delays}
\label{sec:variab}

Most of the results discussed so far are independent of whether the observer 
sees the radiation from the entire reprocessor at the same time, assuming an
illuminating radiation which is constant in time.  In reality, the illuminating 
radiation flux level changes in time, typically being a smoothly decreasing 
function of time. For the ``nearby reprocessor" models 1-5, the finite 
light-travel time differences between different parts of the reprocessor are 
negligible for observer times $t_{obs} \simg 10^2-10^3$ s, particularly if 
the radiation arises from a limited range of solid angles, such as a funnel. 
However, for the ``distant reprocessor" models 6-7, the finite light-travel 
time between the continuum source and the shell means that the observer sees 
simultaneously different parts of the reprocessor which are illuminated
by the continuum at different source times. The regions nearest to the 
observers are illuminated by a continuum corresponding to later source
times than the regions farther from the observer. This convolution can 
be described by an equation of the form
$L(t)={{f}\over{2}} \int_{\theta_{min}}^1{\sin\theta ~d\theta \int_0^{t} dt' 
    ~L_{line}(\theta,t')\delta(t-t'-R(1-\cos\theta)/c))}
={{f}\over{2}} {{c}\over{R}} \int_{max(1,t-2R/c)}^t dt' 
L_{line}(\cos^{-1}(1-{{c}\over{R}}(t-t')),t')$, 
%\eqno{(8)}
%
where $L_{line}(\theta,t)$ is the emitted luminosity from the surface of the 
reprocessor as a function of observing angle and time, $f$ is a factor $\leq$1 
which takes into account the fact that the reprocessor can be clumpy, and 
as seen by the source it can cover less than $4\pi$, and the line emission 
is not isotropic. The integral is over the surface of the reprocessing shell
illuminated by the continuum, which may be beamed (e.g. Weth, \etal 2000).
For distant models the effect of time delays have been considered also by 
Lazzati \etal (1999), B\"ottcher (2000), and in a torus geometry by B\"ottcher 
\& Fryer (2001). The main effect is that the peak line (or reprocessed continuum) 
luminosity is smaller than the line fluxes given in Table 1, due to smearing by 
integration over the surface. For the simplest case where the ionization is
dominated by the initial hard pulse of duration $t_{illum}$ observed at a later
time $t_{obs}$ this would give a dilution factor of the order $t_{illum}/t_{obs}$ 
which could be $\siml 10^{-1}-10^{-2}$, where $t_{obs}\sim (2R/c)(1-\sin\theta_j)
\sim$ day, depending on the model. In such models (e.g. model 7 of table 1) the 
line luminosity actually observed would be diluted below the value $\sim 10^{46}$ 
erg/s given in the table, by an amount which could be substantial, depending on
the luminosity evolution. For instance, if the continuum luminosity $L_{in}\sim 
3 L_x$ of model 7 ($L_{in}=10^{52}$ erg/s, which could not last longer than tens 
of seconds) evolves on a timescale of a day to a value comparable to that of 
model 6 ($L_{in}=10^{47}$ erg/s), and if this spectrum still illuminates the 
shell of gas (i.e. at one day the afterglow shock producing the continuum has 
not outrun the shell, which requires exceptional densities inside the shell, e.g. 
Weth \etal 2000), then the line luminosity would be at least the value 
$10^{44.5}$ given for model 6.  A quantitative discussion of the Fe light curves 
is affected by uncertainties in the model details which would require geometrical 
and parameter space investigations beyond the purposes of this paper. 
However, detailed calculations of specific models (Weth \etal 2000, 
B\"ottcher 2000, B\"ottcher \& Fryer 2001, etc) agree with the above 
approximate estimate of $L_{Fe}\siml 10^{43}-10^{44}$ erg/s if solar 
abundances are used in a ``distant reprocessor" scenario. 

\section{Illumination and Abundance Effects}
\label{sec:abundances}

For the distant models the results are dependent on the illumination model.
If the jet producing the input radiation remains within the reprocessor
shell after 1-1.5 days (requiring rather high intra-shell densities) and
continues to illuminate the reprocessor material as the luminosity evolves 
down to $L_{in}=10^{47} \sim 3 L_x$ erg/s in model 6, then 30xsolar abundances 
are sufficient.  
However if the afterglow shock or jet producing the illuminating continuum 
outruns the reprocessor shell in less than a day, then the effective 
continuum is a prompt flash of duration much less than a day with a
luminosity comparable to model 7, but the line intensities of model 7 would
be affected by a dilution factor $10^{-2}-10^{-3}$ due to the time delay 
smearing discussed in the previous section, and larger increases in the
solar abundance relative to solar may be needed.

Relevant to the abundance issue, e.g. in the distant reprocessor scenarios
if these are associated with recent supernova events (e.g. Piro et al 2000, 
Vietri et al 2000, Reeves et al 2002), is the production of adequate amounts 
of iron via the decay of nickel and cobalt, involving a delay of order 70 days. 
If the supernova occurred less than a few months before the GRB, the shell
may contain supersolar nickel, but not much iron.  In nearby models, if 
supersolar abundances are required, highly nickel enriched material may be 
entrained by the jet from the core to the outer edges of the funnel in the 
envelope. In the nearby models, the case has been made that multiple Compton
scatterings in the stellar funnel (McLaughlin et al 2001) will cause the
nickel line energies to mimic those of iron. In the distant models, 
multiple scatterings are not expected, so the necessity for producing
iron is harder to avoid. In our models, we can crudely test for the 
effects of having nickel instead of iron, by modifying the abundances 
in the model scenario which most nearly resembles the supernova reprocessor.  
We have done this in model 6ni, in which we have used the conditions for 
model 6 but we have set the iron abundance to 0 and instead chosen a nickel 
abundance such that the number density of nickel ions is the same as the 
number density of iron ions in model 6.  This corresponds to a 20 times 
overabundance of nickel relative to the solar values of Grevesse et al (1996).
(Owing to uncertainties in atomic data, xstar does not include cobalt, 
so we cannot directly test scenarios involving mixtures of this element). 
The results are given in Table 1, in which the line strengths in the "Fe"
column for model 6ni correspond to nickel rather than iron.  We find that 
the dominant nickel line is the helium-like complex at 7.78 keV, and the 
strength of this feature n model 6ni somewhat exceeds the strength of the 
Fe XXVI line in model 6.  This feature appears prominently in the model 
spectrum, and would lead to a greater inferred redshift for the source, 
if it were the true origin of the feature observed in eg. GRB991216. 

\section{Discussion}
\label{sec:disc}

The results of the previous sections indicate that iron line luminosities 
in the range of $\sim 10^{43} - 10^{45.5}$ erg s$^{-1}$, comparable to the 
luminosities observed so far (eg. Piro et al. 2000), can be produced by 
photoionization of a dense reprocessing gas in the vicinity of gamma ray 
burst sources.  The model densities and ionization parameters assumed in 
both ``nearby" scenarios (e.g. the jet plus bubble model of \Mesz \& Rees 
2001 or the delayed jet model of Rees \& \Mesz, 2000), and in ``distant" 
scenarios (e.g. the pre-existing supernova shell of Lazzati et al 1999, 
Piro et al 2000, Vietri et al 2001), are generally able to do this, with
higher line luminosities achieved if one uses Fe abundances $\sim 30-100$ 
times solar and/or large incidence angles $\theta_i \simg 80^o$. 
However, the models so far are highly simplified, and while they address
particular aspects, they are not yet at the stage where they can provide 
a general fit to all the properties of particular afterglows where X-ray
lines have been reported. This is partly due to the small and sparse set 
of line observations, whose significance level is not high, and to the complex 
nature of the models, involving a number of poorly constrained parameters.

An Fe-group metal overabundance relative to solar is plausible both in nearby 
models (where the stellar progenitor funnel walls can be enriched by the jet or 
bubble bringing up enriched core material) and in the distant model, where a 
pre-existing supernova shell would also consist of core-enriched material.
However, in both nearby and distant models (in the latter if the shell age is 
less than $\sim 70$ days) the heavy ions may be mainly Ni instead of Fe.
In nearby models there is a plausible way of degrading Ni lines to resemble
Fe lines through multiple scattering (McLaughlin et al 2001), while in distant
SN shell models a somewhat older shell is required, or some other mechanism 
for making Ni appear as Fe.

Another important parameter in photoionization models of line formation 
is the incidence angle at which the input continuum reaches the 
reprocessing gas. This, at least for the simple models considered here, 
is subject to some natural constraints. For nearby models involving 
a funnel in a stellar envelope illuminated by a jet or bubble, a large 
incidence angle is quite plausible. On the other hand for distant models, 
such as a supernova  shell ejected days or months before the burst, the 
radiation is unlikely to reach the shell at a large incidence angle. Even if 
the shell is lumpy, the incidence angle would be expected to be closer to normal.

A significant observational constraint on the models is the line to continuum 
ratio (Lazzati, Ramirez-Ruiz and Rees, 2002; Ghisellini, Lazzati, Rossi \& 
Rees 2002). A nominal ``target" value is given by the Chandra observations of 
GRB 991216 (Piro et al 2000), for which an equivalent width EW$=0.5 \pm 0.013$
keV is quoted in discussing the observer frame spectrum.
%This can be loosely translated into a line/continuum ratio 
%$F_{line}/F_{cont} =  (EW/E_{line})\sim 0.15$.
The observer frame energy equivalent width is related to the source frame 
value by EW$_{ob}$=EW$_{em} 1/(1+z)$, and the object is at $z\sim 1$, so 
the GRB 991216 source frame EW $ \sim 1$ keV is the quantity to be compared 
against our source frame calculations. In Table 1 we have defined the 
line/continuum ratio based on the continuum in the 1-10 keV range, hence 
using $E_{max}=10$ keV, $E_{min}=1$ keV and a canonical -1 spectrum the 
relation is EW$ = (F_{line}/F_{cont}) E_{line} \ln(E_{max}/E_{min})$, or
$(F_{line}/F_{cont})_{GRB991216}=({\rm EW}/E_{line})/ln(E_{max}/E_{min})\sim 0.06$. 
This measure of line reprocessing efficiency depends on the 
continuum measured over a wide energy range, and  
in the objects with line detections so far many of these photons are 
redshifted out of the observable energy band.
%However, the upper and lower energy boundaries chosen are instrument 
%and choice dependent, so for simplicity below we will use as a nominal 
%GRB 991216 source frame value a simpler estimate, which avoids energy band 
%and spectral considerations, $F_{line}/F_{cont} =  (EW/E_{line}) \sim 0.04$. 

Table 1 shows that the models that approach the
GRB 991216 comparison value for the line continuum ratio of 0.06 
(or -1.2 in log scale) are Model 2"x100 
(nearby, 87$^o$ incidence angle, 100xsolar) which has a line/continuum ratio 
$\sim 0.02$, and Model 6''x30 (distant, 87$^o$ incidence angle, 30xsolar),
which has a line/continuum ratio of $\sim 0.1$.  The GRB 991216 comparison value 
for the line equivalent width of 1 keV (in the source frame) is achieved or surpassed 
by 2''x30, 2''x100, 6'' and 6''x30.
Equivalent width is a measure of reprocessing efficiency which depends on the 
continuum in the immediate vicinity of the line.
The equivalent widths we calculate approach or exceed 0.5 keV for the nearby and 
distant models which include both non-normal incidence and an iron over-abundance.  

 For the same high
incidence angle and a lower overabundance, the distant model makes a stronger 
line than the nearby model. This has been explained in \S \ref{sec:comparison} 
in terms of the shallower line depths in high incidence angle cases, with the
lower density of distant models leading to more penetration of Fe line photons.

However, the line/continuum model values of Table 1 were calculated in the 
spirit of investigating how the physics of the line production varies as a
function of the basic model parameters. In particular, the abundances and the 
incidence angles were varied more or less arbitrarily, and this needs to be 
supplemented with astrophysical considerations of how plausible particular 
parameter values are in the context of given models. 

In distant scenarios a large Fe overabundance is reasonable, if there is a
weeks to months delay between the SN and GRB explosions, requiring good 
timing so that enough Fe has been formed but the shell has neither dispersed 
nor is too close. In nearby scenarios, a large Fe group overabundance is 
also plausible, as matter is dredged up from the core by the jet. Furthermore, 
multiply down-scattered Ni lines, as they bounce in the funnel, can mimic 
the Fe lines (McLaughlin et al 2002). This possibility of Ni mimicking Fe does
not work in distant scenarios, where multiple scattering are not expected. 
Another aspect of multiple reflections in a  stellar funnel is that it can 
increase somewhat further the line to continuum ratio (Ghisellini et al 2002).
Other possible difficulties have been pointed out for nearby  models, e.g. 
Ghisellini et al 2002, which were based on analytical calculations for
normal incidence conditions.

A large incidence angle is naturally expected in nearby models, from the
geometry of a funnel in a stellar envelope. An example of this is the 
nearby model 2''x100 with 87$^o$ incidence angle and 100xsolar Fe, which 
has a line/continuum ratio of 0.02 and an equivalent width of 1.28 keV. 
On the other hand, a large incidence 
angle is less likely in a distant supernova shell model, where quasi-normal 
incidence is the natural expectation. A normal incidence distant model such 
as 6x100, even with 100xsolar Fe, produces a line/continuum ratio of 0.001 
and an equivalent width of 0.011 keV (Table 1 and Figure 5b). 
Other models, in general, have lower values.

Both nearby and distant models, in their simple versions as discussed here 
and elsewhere, are constrained by total energetics (Ghisellini etal 2002, 
Kumar \& Narayan, 2002), and these issues were not addressed here. As far as
the ability of these models to reproduce the observed nominal 0.06 line/continuum 
ratios or 1 keV equivalent widths, values approaching this can be 
achieved in distant (supernova) models 
if a large incidence angle is used, which for this model appears implausible. 
Values within a factor 3 of this can be achieved with nearby (stellar funnel) 
models, using optimistic but plausible parameters.
The line detection significance in this object is $4.5\sigma$ for K$\alpha$ and 
K-edge identification, or $3.5\sigma$ for the K$\alpha$ alone (Piro et al 2000). 
Given the 3-4$\sigma$ confidence level in the existence of the lines and the
line/continuum ratios and equivalent widths, as well as the 
highly approximate nature of the models, 
it may be too early to reach strong conclusions about preferring one model 
over another.

The exact shape of the line, and the fraction escaping in a narrow core, 
are sensitive to the geometry and other model details of the reprocessor. 
For this reason, we have not attempted here to make a detailed comparison
of models to specific observations, but rather concentrate on the more 
general question of the effect on observable quantities of various physical
properties inherent in the two main generic classes of models which have
been recently discussed. In the two scenarios the line is computed based on 
the continua listed in Table 1. For the ``nearby" scenarios, we can
assume that the continuum observed is $\simg$ the continuum exciting the 
observed lines. As emphasized by Lazzati etal 2002, this is an upper limit, 
since a fraction of the continuum may reach the observer directly. The 
X-ray lines are generally seen accompanied by a bump-like rise in the X-ray 
continuum, which in the nearby scenario is attributed to a rising bubble or 
a secondary late jet component producing its own X-ray power law which 
interacts with the outer stellar envelope (Rees \& \Mesz 2000, \Mesz \& 
Rees 2001). The line timescale of 1-2 days is caused by the intrinsic 
timescale of the (sub-relativistic) bubble rise or secondary long-term jet, 
and relativistic or geometrical time delays are negligible, since the 
illumination timescale is comparable to the observation time. This component 
is somewhat in excess of that attributed to the canonical relativistic main 
jet, which is thought to produce the GRB at early times, and which at 
$t \simg 1$ day when its Lorentz factor has dropped to values $\sim$ 10 at 
radii $\simg 10^{16}$ cm is thought to give rise to the standard power law 
(but not to X-ray lines) seen in canonical X-ray afterglows.

In ``distant reprocessor" scenarios involving, e.g. invoking a supernova 
shell (Lazzati et al 1999, Vietri et al 2001, Piro et al, 2000, Weth et al, 
2000, B\"ottcher, 2000, B\"ottcher and Fryer 2001, Ballantyne \& Ramirez-Ruiz 
2001), the photoionizing continuum is assumed to be due to the same canonical
jet  responsible for the GRB. The shell distance in this model is 
determined from geometrical considerations (time delay $\sim (R/c)(1-\cos
\theta_{sh})\sim 1-2$ days, where $\theta_{sh}$ is shell effective angle).
For such distant models, unless the density inside  the shell exceeds 
$\sim 10^6-10^7$ cm$^{-3}$ the jet producing the X-ray continuum would
have moved beyond the shell at $t\sim 1$ day, and hence the appropriate
photoionizing luminosity is that at early times while the jet is inside 
(e.g. model 7); the line luminosities in Table 1 for model 7 have to be 
multiplied by a time-delay dilution factor which would be $\sim t_{ill}/t_{obs} 
\siml 10^{-1}-10^{-2}$, and this might make the line to continuum ratio lower
than observed. More detailed time delays have been discussed in the literature 
cited, using however specific model geometries, which is not our purpose here. 
On the other hand, within the spirit of our approximate geometry models,
the distant model 6 (in particular the 30xtimes solar Fe model 6x30) at
high incidence angles has an input luminosity corresponding to a time delay 
factor $t_{ill}/t_{obs}\siml 1$, and leads to line to X-ray continuum
ratios in the observed range. Two additional assumptions involved in distant 
scenarios are a) that line-producing ejecta shell has had months to decay 
from Ni into Fe, requiring a supernova to have occurred months before the 
burst, for which stellar evolutionary scenarios are currently very speculative; 
and b) that the shell at $R\sim 10^{16}$ cm of total mass $\sim$ few solar 
masses is either geometrically very thin, $\Delta R/R\siml 10^{-3}$, or else 
it consists of very dense blobs whose density happens to provide a covering 
factor of order unity.

The recent reports of XMM observations of Mg, Si, S lines but no Fe lines
from GRB 011211 (Reeves etal 2002) and from GRB 020813 (Butler, et al 2003),
which appeared months after submission of this paper, is in strong contrast 
to the 5 previous GRB line detections 
referred to elsewhere in this paper (e.g. Piro et al 2000, etc). If this 
interpretation is correct, it would imply different conditions in this burst
compared to the previous five bursts which showed Fe lines. E.g. in 
photo-ionized models, this might arise from a different ionizing spectrum, 
ionization parameter or illumination history, etc. (Lazzati, et al, 2002).
We note that the collisional ionization mechanism favored by Reeves 
et al (2001) would imply very high densities for a SN shell ($10^{15}$ /c.c), 
and such densities would be more naturally expected in nearby reprocessor 
scenarios, in addition to conditions which could lead to mixed collisional 
ionization and photoionization. 

Further diagnostic information about the reprocessor is available from 
detections or limits on absorption features due to bound-bound or bound-free 
transitions of iron.  For the physical conditions envisaged in this paper, 
photoionization equilibrium is a good approximation, and non-equilibrium
effects are expected to be small. With different model assumptions, however,
these might play a role, e.g possibly in enhancing the radiation recombination
(free-bound) edge (Yonetoku, \etal, 2001), Yoshida \etal, 2001), or in 
producing an absorbing column which varies with time after the initial burst 
onset (Lazzati, Perna \& Ghisellini 2001; Lazzati \& Perna 2002). 
Enhanced recombination may occur if the 
electron temperature is (very) low compared with the ionization temperature,
which is less likely under photoionization conditions, but may be possible 
if the gas undergoes sudden rarefaction and adiabatic cooling of the electrons. 
The data on GRB 991216 which they discuss is close to what is predicted by 
the simple photoionization equilibrium models discussed here. 
In general, the bound-free absorption cross section from the K shell 
of iron is not a sensitive function of the ionization state of iron, 
and it is comparable with the Thompson cross section if the abundances are 
cosmic and if the ionization is favorable.  Since all the reprocessors 
described in the previous section are effectively semi-infinite, they will 
not transmit efficiently near iron and so absorption features (e.g. as
reported, at the 3 $\sigma$ level, by Amati \etal, 2000) are not expected 
from these simplified models.  
Absorption features would be imprinted on the reflected continuum from thick 
reprocessors at ionization parameters lower than those we examine here, e.g. 
log($\xi) \leq$ 100, but this does not appear to be compatible with conditions 
inferred from observed emission lines.  However, if the Thompson depth of the 
reprocessor were, at least temporarily, close to unity (as might be expected 
in a nearby reprocessor model, as the jet and the prompt portion of its 
relativistic waste bubble breaks through the last few optical depths at 
increasing angles), such features may also be naturally expected. 

In conclusion, we have investigated both nearby and distant models of GRB 
afterglow reprocessor geometries proposed as sources for the reported X-ray 
lines in several GRB, through photoionization by an incident continuum. 
We find that the effects of Comptonization and pair formation can affect the 
results, depending on the conditions assumed.  The absolute values of the 
line luminosities can be reproduced fairly well by both models, if
high overabundances and high incidence angles are assumed. While distant 
(e.g. supernova shell) models are more effective at producing high 
line/continuum ratios and equivalent widths at high incidence angles, such angles are not
expected to occur naturally without some further assumptions being 
introduced, and at quasi-normal incidence the line ratios are too low.
Nearby (e.g. stellar funnel) can account naturally for high incidence angles, 
and reach line/continuum ratios and equivalent widths within a factor 3 of the reported values.
Issues remain concerning overall agreement with the entire burst history as 
well as energetics.  Further line observations at higher significance levels, 
as well as more detailed modeling, will be required before strong conclusions 
can be reached concerning the type of progenitors and geometries involved.

\acknowledgments
We are grateful to Xin-min Hua for use of the Comptonization code, 
D. Ballantyne, E. Ramirez-Ruiz and D. Lazzati for discussions, and
to NASA NAG5-9192 and the Royal Society for support.

\newpage

\references

Amati, L, \etal, 2000, Science, 290, 953.

Ballantyne, D., and Ramirez-Ruiz, 2001 Ap.J. Lett. 559, 83

Ballantyne, D., Ross, R., and Fabian, A., 2001 MNRAS 327, 10.

Bautista, M., $\&$ Kallman, T. 2001, ApJ 134, 139

B\"ottcher, M , 2000, ApJ 539, 102

B\"ottcher, M and Fryer, C, 2001, ApJ 547, 338

Bussard, R., Ramaty, R., and Drachman, R., 1979 ApJ 228 928

Butler, R.N. et al, ApJ submitted (astro-ph/030353)

Coppi, P., and Blandford, R., 1990 MNRAS 245 453

Ghisellini G, Lazzati D, Rossi E \& Rees M.J. 2002, Astron. Ap., 389, L33 

Grevesse, N., Noels, A., and Sauval, A., 1996, in  
``Cosmic Abundances'' ASP Conference Series, 99, S. Holt and G. Sonneborn, eds.

Guilbert, P., 1986 MNRAS 218 171

Hua, X., 1997 Comp Phys 11, 660

Kallman, T., and Bautista, M., 2001, ApJ 133, 221

Kallman, T.,  and McCray, R., 1982, Ap. J. Supp., 50, 49

Krolik, J., McKee, C., and Tarter, C. B.,1981 Ap. J. 

Kumar, P \& Narayan, R, 2002, ApJ subm., astro-ph/0205488

Lazzati, D. \etal (1999) MNRAS, 304, L31

Lazzati, D. \& Perna, R. 2002, M.N.R.A.S. 330, 383

Lazzati, D., Perna, R. \& Ghisellini G, 2001, M.N.R.A.S. 325, L19.

Lazzati, D., Ramirez-Ruiz E and Rees, M.J., 2002, astro-ph/0204319
 
McCray, R., Wright, C., and Hatchett, S., 1978, Ap. J.

McLaughlin, G.C., Wijers, R., Brown, G, \& Bethe, H., 2002, Ap. J. 567, 454

\Mesz, P., and Rees, M., 2001, ApJ(Letters), 556, L37 

Nayakshin, S., et al., 2000 ApJ 537 833

Piro, L., et al, 1999, Ap J (Letters), 514, L73

Piro, L., et al, 2000 Science 290 955

Rees, M.J. and \Mesz, P, 2000, ApJ (Letters), 545, L73

Reeves, J.N. et al., 2002, Nature 416, 512

Tarter, C. B., Tucker, W,  $\&$ Salpeter, E.  1969, ApJ, 156, 943

van Paradijs J, Kouveliotou C \& Wijers R.A.A.J., 2000, A.R.A.A. 38, 379

Vietri, M., \etal, 2001 ApJ (Letters), 550 L43

Weth, C., \etal, 2000, ApJ 534, 581

Yonetoku, D, \etal, 2001, ApJ, 557, L23

Yoshida, A, \etal, 1999,  Astron. Ap., 138, 433

Yoshida, A, \etal, 2001, ApJ, 557, L27

\clearpage

\begin{figure}[!bp]
\plotfiddle{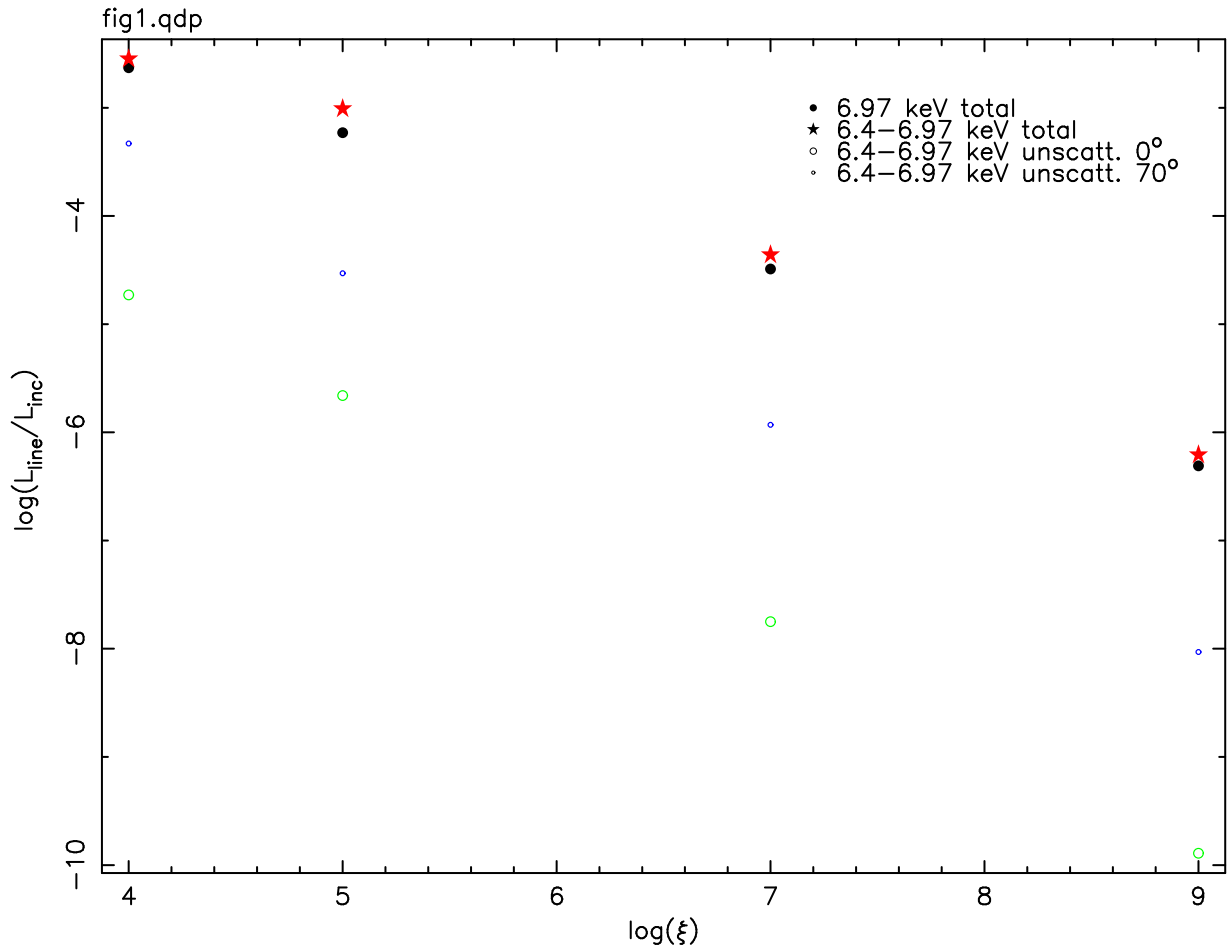}{18cm}{270}{70}{70}{-260}{420}
\caption{Line reprocessing efficiency vs. ionization parameter
for the nearby models shown in the table.}
\end{figure}

\clearpage

\begin{figure}[!bp]
\plotfiddle{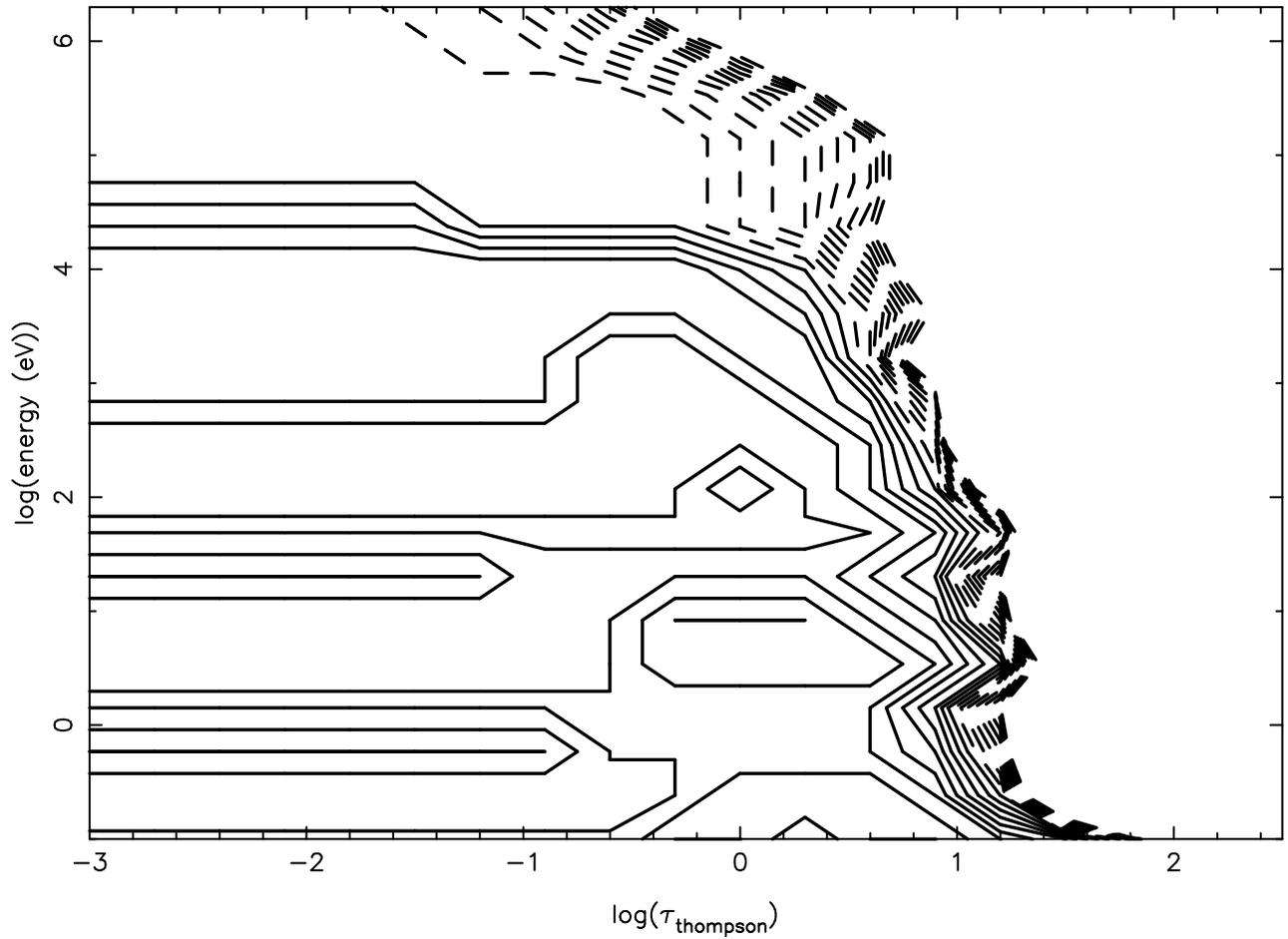}{16cm}{270}{70}{70}{-260}{420}
\caption{ (a) Contours of constant mean intensity relative to 
the incident intensity  vs. energy and Thomson
optical depth for model 2.  Contours interval is 0.2 dex.  Solid contours 
correspond to mean intensity greater than incident and domonstrate the effects
of Compton downscattering.  Dashed contours correspond to mean intensity 
less than incident and demonstrate the effects of attenuation and 
Compton reflection. }
\end{figure}

\clearpage

\setcounter{figure}{1}

\begin{figure}[!bp]
\caption{ (b) Temperature vs. depth for model 2.}
\plotfiddle{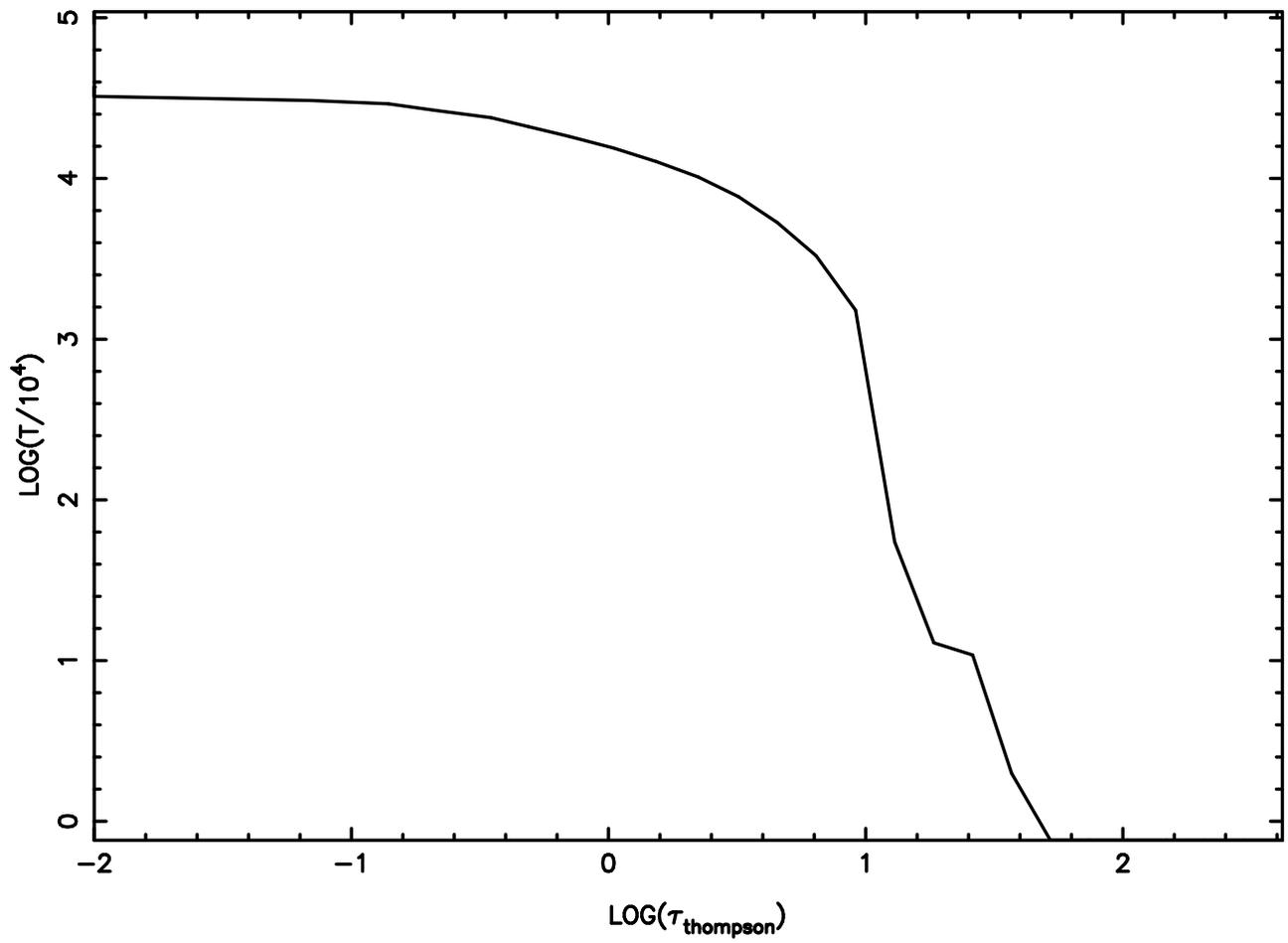}{18cm}{270}{70}{70}{-260}{420}
\end{figure}

\clearpage

\setcounter{figure}{1}

\begin{figure}[!bp]
\plotfiddle{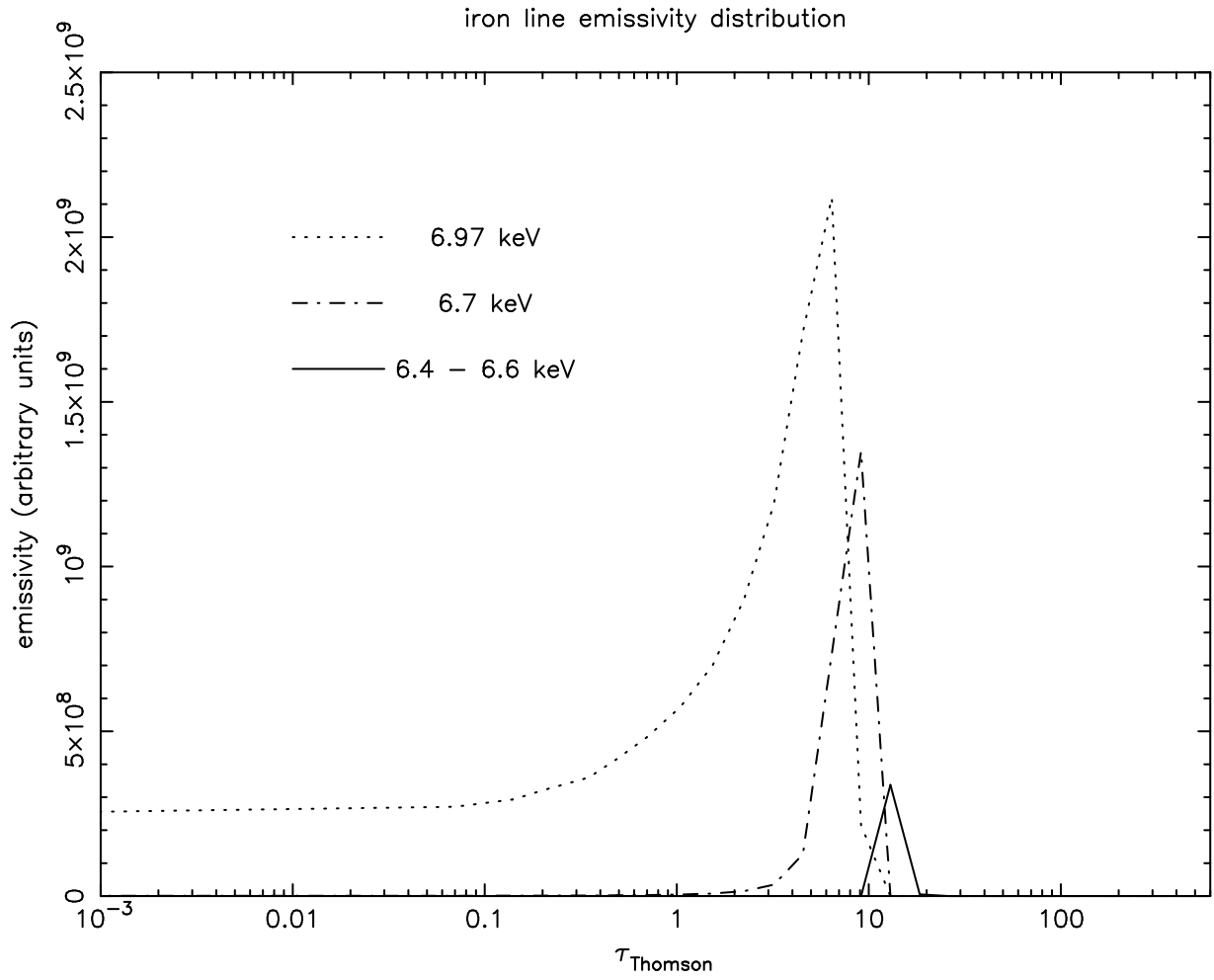}{18cm}{270}{70}{70}{-260}{420}
\caption{ (c) Iron line emissivity vs. depth for model 2 for the 
various components:  6.97 keV (hydrogen-like iron),
6.7 keV (He-like iron), and 6.4 -- 6.6 keV (all lower stages).}
\end{figure}

\clearpage

\begin{figure}[!bp]
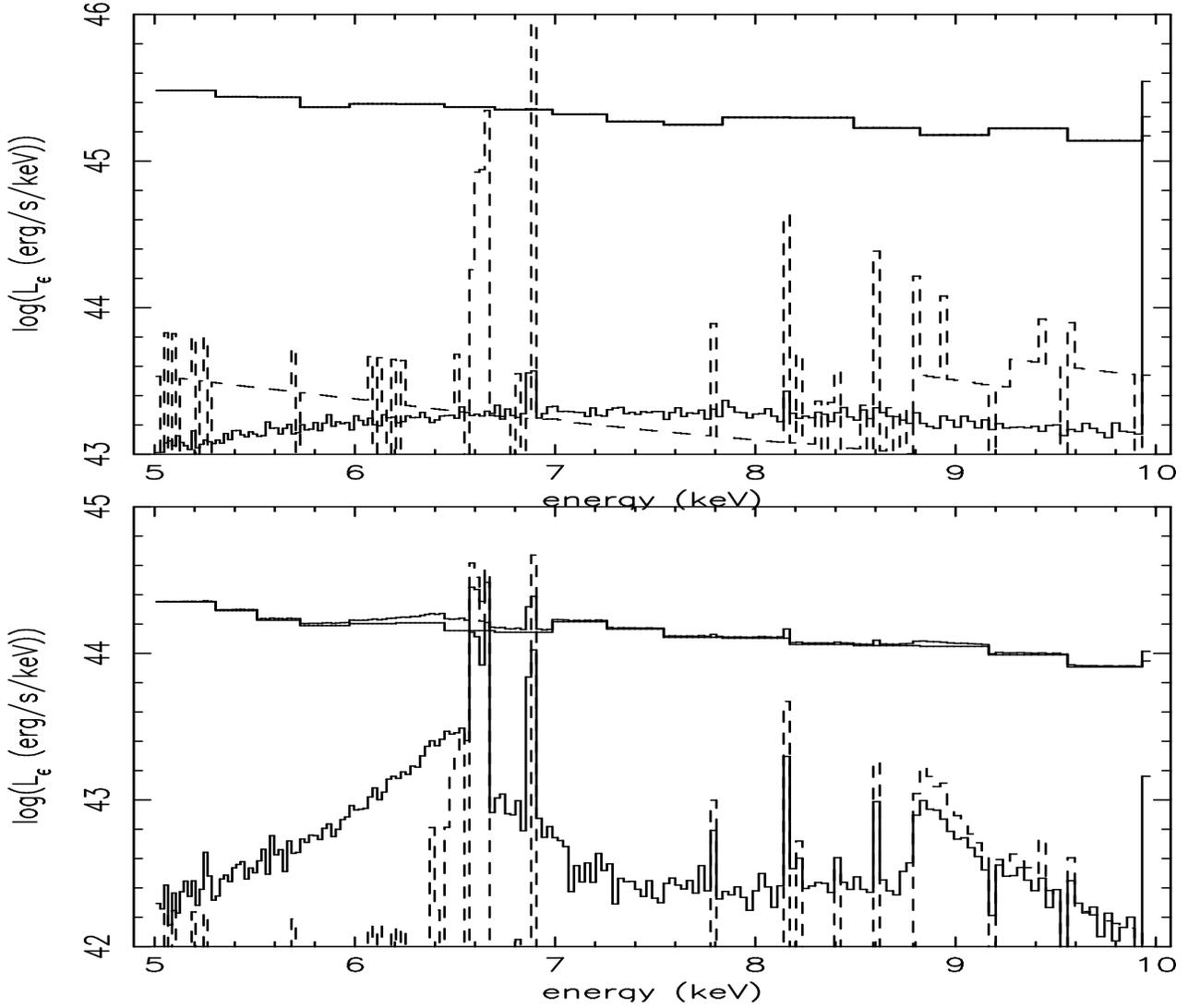

\plotfiddle{f3a.ps}{6.5cm}{270}{70}{40}{-260}{220}
\plotfiddle{f3b.ps}{6.5cm}{270}{70}{40}{-260}{220}
\caption{Reflected (solid) and emitted (dashed) rest frame  spectra from nearby models with 
solar abundance of iron, varying the incidence angle. Units are specific luminosity, 
corresponding to the reflected and reprocessed spectrum from gas surrounding 
a source with ionizing luminosity 10$^{47}$ erg/s.
Solid curves correspond to Comptonized  thermal emission, 
Compton reflected continuum, and total. 
Panel a) Model 2 with normal incidence.
Panel b) Model 2'' with non-normal incidence ($\mu$=0.05).}
\end{figure}

\clearpage

\begin{figure}[!bp]
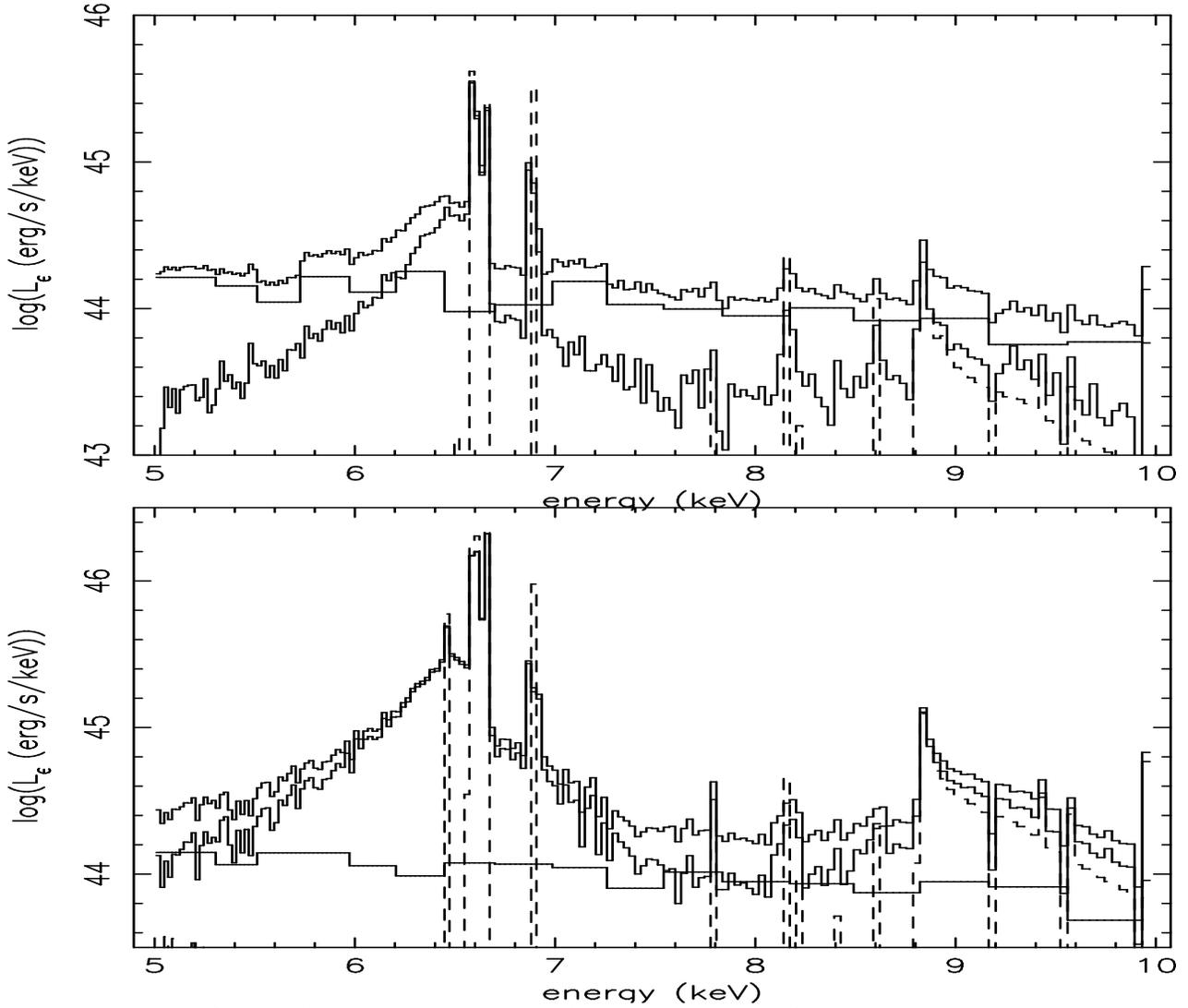

\plotfiddle{f4a.ps}{6.5cm}{270}{70}{40}{-260}{220}
\plotfiddle{f4b.ps}{6.5cm}{270}{70}{40}{-260}{220}
%\plotfiddle{f4c.ps}{6.5cm}{270}{70}{40}{-260}{220}
\caption{ Reflected (solid) and emitted (dashed) spectra from nearby models 
with non-normal incidence and enhanced iron.  Units are specific 
luminosity, corresponding to the reflected and reprocessed spectrum from gas 
surrounding a source with ionizing luminosity 10$^{47}$ erg/s.  
Solid curves correspond to Comptonized  thermal emission, 
Compton reflected continuum, and total. 
Panel a) Model 2'' with $\mu=0.05$ and Fe=30x solar.
Panel b) Model 2'' with $\mu=0.05$ and Fe=100x solar.  
%Panel c) Model6'' with Fe=30x solar.
}
\end{figure}

\clearpage

\begin{figure}[!bp]
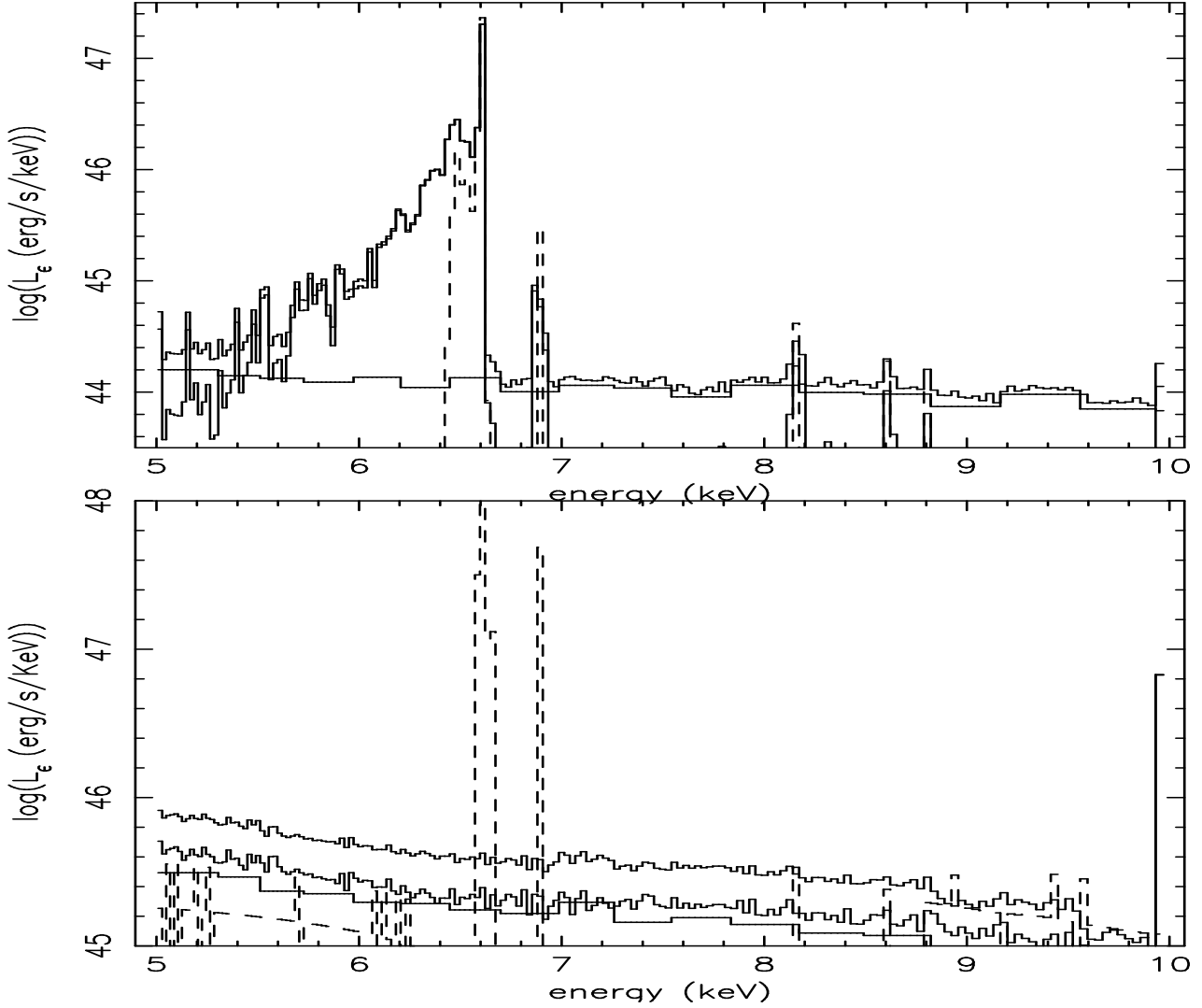

\plotfiddle{f5a.ps}{6.5cm}{270}{70}{40}{-260}{220}
\plotfiddle{f5b.ps}{6.5cm}{270}{70}{40}{-260}{220}
%\plotfiddle{f4c.ps}{6.5cm}{270}{70}{40}{-260}{220}
\caption{ Reflected (solid) and emitted (dashed) spectra from distant models 
with enhanced iron, varying the incidence angle. 
Units are specific luminosity, corresponding to the reflected and reprocessed 
spectrum from gas surrounding a source with ionizing luminosity 10$^{47}$ erg/s.
Solid curves correspond to Comptonized  thermal emission,
Compton reflected continuum, and total.
Panel a) Model 6''x30, $\mu=0.05$ with Fe=30x solar.
Panel b) Model 6x100,  normal incidence with Fe=100x solar.
}
\end{figure}

\clearpage

\end{document}